\documentclass[journal]{IEEEtran}
\usepackage{amsmath,amsthm,amsfonts,amscd,amssymb,graphics,dsfont}
\usepackage[mathscr]{eucal}
\usepackage{graphics,graphicx,multicol}
\usepackage{epsfig}
\usepackage{epstopdf}
\usepackage{subfigure}
\usepackage{stfloats}
\usepackage{latexsym}
\usepackage{amsfonts}
\usepackage{cite}
\usepackage{algorithm,algorithmic}
\usepackage{color}

\newtheorem{rem}{Remark}

\begin{document}
%%%%%%%%%%%%%%%%%%%%%%%%%%%%%%%%%%%%%%%%%%%%%%%%%%%%%%%%%%%%%%%%%%%%%%%%%%%%%%%%%%%%%%%%%%%%%%% Title
\title{A Deep Learning Approach to Universal Binary Visible Light Communication Transceiver}
\author{Hoon Lee,~\IEEEmembership{Member,~IEEE}, Tony Q. S. Quek,~\IEEEmembership{Fellow,~IEEE}, and Sang Hyun Lee,~\IEEEmembership{Member,~IEEE}
\thanks{This work was supported in part by the National Research Foundation of Korea (NRF) grant
funded by the Korea government (MSIT) (No. 2019R1F1A1060648,2019R1A2C1084855), in part by a Korea University Grant, and in part by the SUTD Growth Plan Grant for AI and in part by the SUTD-ZJU Research Collaboration under Grant SUTD-ZJU/RES/05/2016. This paper will be presented in part at IEEE Global Communications Conference Workshop 2019,
HI, USA, Dec. 2019 \cite{HLee:19b}. {\em (Corresponding
author: Sang Hyun Lee.)}

H. Lee is with the Department of Information and Communications Engineering, Pukyong National University, Busan 48513, South Korea (e-mail: hlee@pknu.ac.kr). T. Q. S. Quek is with the Singapore University of Technology and Design, Singapore 487372, and also with the Department of Electronic Engineering, Kyung Hee University, Yongin 17104, South Korea (e-mail: tonyquek@sutd.edu.sg). S. H. Lee is with the School of Electrical Engineering, Korea University, Seoul 02841, South Korea (e-mail: sanghyunlee@korea.ac.kr).

\copyright 2018 IEEE. Personal use of this material is permitted. Permission from
IEEE must be obtained for all other uses, in any current or future media,
including reprinting/republishing this material for advertising or promotional
purposes, creating new collective works, for resale or redistribution to servers
or lists, or reuse of any copyrighted component of this work in other works.}
}\maketitle %\thispagestyle{empty}
%%%%%%%%%%%%%%%%%%%%%%%%%%%%%%%%%%%%%%%%%%%%%%%%%%%%%%%%%%%%%%%%%%%%%%%%%%%%%%%%%%%%%%%%%%%%%%%%%%%%%%%%AbstractX
\begin{abstract}
This paper studies a deep learning (DL) framework for the design of binary modulated visible light communication (VLC) transceiver with universal dimming support. The dimming control for the optical binary signal boils down to a combinatorial codebook design so that the average Hamming weight of binary codewords matches with arbitrary dimming target. An unsupervised DL technique is employed for obtaining a neural network to replace the encoder-decoder pair that recovers the message from the optically transmitted signal. In such a task, a novel stochastic binarization method is developed to generate the set of binary codewords from continuous-valued neural network outputs. For universal support of arbitrary dimming target, the DL-based VLC transceiver is trained with multiple dimming constraints, which turns out to be a constrained training optimization that is very challenging to handle with existing DL methods. We develop a new training algorithm that addresses the dimming constraints through a dual formulation of the optimization. Based on the developed algorithm, the resulting VLC transceiver can be optimized via the end-to-end training procedure. Numerical results verify that the proposed codebook outperforms theoretically best constant weight codebooks under various VLC setups.
\end{abstract}
\begin{IEEEkeywords}
Visible light communication, deep learning, dimming support, primal-dual method
\end{IEEEkeywords}

\section{Introduction}
Visible light communication (VLC) recently emerges as a promising alternative to complement over-crowded radio frequency (RF) bandwidth communication systems for broadband data transmission \cite{Grubor:08,HElgala:11,SHLee:15}. Several key advantages offered in comparison to existing RF systems, such as license-free operation, high efficiency, low cost, high bandwidth, high spatial diversity, electromagnetic interference free channel, and controlled beam shaping, etc., notch solid-state lighting devices realized by light emitting diodes (LEDs) the most essential component for VLC, providing high-quality illumination and high-capacity communication simultaneously \cite{Rajbhandari:17,HElgala:11,SHLee:15,SLi:16}. Besides communication, LED-based VLC has a number of emerging applications including positioning, sensing, object ranging and detecting \cite{PPathak:15,JLuo:17}.

Intensity modulation and direct detection (IM/DD) technique is considered a major approach to convey a message via a nonnegative real-valued optical signal \cite{KAhn:12,SHLee:13,XLiang:17}. In addition, the average and peak intensity restriction imposed by LED property, human physiologic effect and external preference require a careful design of binary transmission strategy when dimming control is achieved by utilizing on-off keying (OOK) \cite{SKim:11,SHLee:12,SZhao:16,SZhao:17}. However, existing techniques lack practical considerations about channel imperfection, signal-dependent noise source, etc.

\subsection{Related Works and Motivation}
Deep learning (DL) techniques recently have been applied for the optimization of end-to-end physical layer communication systems \cite{OShea:17,Dorner:18,MKim:18,BKaranov:18,HLee:18a,HLee:18b,HLee:19}. The use of deep neural networks (DNNs) for joint encoding and decoding in RF communication is first introduced in \cite{OShea:17}. In particular, a machine learning framework called an autoencoder (AE) \cite{YBengio:09} has been employed to design a pair of wireless transmitter and receiver over fading channels via a single DL training process \cite{OShea:17}. This result has been extended successfully to various scenarios such as orthogonal frequency division multiplexing (OFDM) \cite{MKim:18} and optical fiber communication \cite{BKaranov:18}.

The recent works in \cite{HLee:18a,HLee:18b,HLee:19} investigate the feasibility of the AE framework in the dimmable VLC system design. Compared to the RF counterparts in \cite{OShea:17,Dorner:18,MKim:18}, the DL-based VLC optimization requires to control the behavior of DNNs by including optical signal constraints, e.g., intensity nonnegativity and dimming support, into the training process. A multi-colored VLC scenario is considered in \cite{HLee:18a} where red/green/blue LEDs are utilized for the message transmission. A postprocessing method based on a projection operation is employed such that the hidden output of the DNN satisfies the exact target dimming constraint. In \cite{HLee:18b} and \cite{HLee:19}, a DNN transceiver of the OOK-modulated VLC subject to a nonconvex binary constraint is designed to generate the set of feasible OOK pulses. A deterministic binarization technique is developed by {\em annealing} the continuous-valued DNN output gradually to the binary vector during multiple stages of the DNN training.
%A careful design of VLC transmission system involves numerically controlled constraints, which can be handled by a state-of-the-art DL technique.

Those previous works prove very efficient for handling optical impairment issues such as signal-dependent noise and receiver filter imperfection \cite{HLee:18a,HLee:18b,HLee:19}. However, such works adopt DL design techniques of the VLC transmitter and receiver that work for a specific dimming target, and the adaptation to arbitrary dimming requirements lacks. To make the resulting transceiver pair practical, the design technique necessarily considers the support for all possible dimming requirement candidates. However, this goal incurs a very challenging task of combining multiple transceiver pairs developed for distinct requirements into a single universal one, in that a distinct dimming requirement is associated with the corresponding mathematical formulation of the constraint, and the resulting individual transceiver structures may be disparate. The optimized transceiver is required to operate universally subject to the adaptive satisfaction of all different constraints. Specifically, machine learning techniques for a digital system involve a training task of discrete output generation, which is known to be very challenging for vanishing gradient issues. Previous works \cite{HLee:18a} use projection and multistage annealing techniques to handle those issues for a single dimming requirement. In the case of multiple dimming support, an intuitive strategy for simultaneous training could be the successive application of such techniques to mini-batches associated with individual dimming requirements. However, such a strategy is readily found intractable since the parameter updates for driving the convergence of a universal neural network are unwieldy and the overall computational efforts for simultaneous training are excessive. Therefore, a new machine learning approach to handle multiple dimming requirements and the discrete output message set is necessary.

\subsection{Contributions and Organization}
This work investigates a DL framework for the design of dimmable OOK-modulated VLC systems working {\em universally} under arbitrary dimming requirement. %For universal support of arbitrary dimming requirement, the average SER minimization problem is subject to simultaneous satisfaction of multiple dimming constraints.
To this end, a generalization of the dimming condition \cite{SZhao:17} is considered such that the optical intensity of the transmitted optical pulse is controlled by the average number of ones over the OOK codebook. Such a binary codebook design approach has been shown to provide an improved spectral efficiency performance as compared to classical constant weight codes (CWCs) \cite{SZhao:16,Ostergard:10}. However, %in terms of the decoding error rate performance,
it is an involved task to identify an efficient OOK codebook with generalized dimming targets and the corresponding decoding strategy over noisy channels, since it, in general, requires computationally expensive search over all possible binary codebooks. To handle this challenge, a DNN-based VLC transceiver structure comprised of encoding and decoding neural networks is proposed %so that the modulation and demodulation of the end-to-end VLC system are addressed by encoding and decoding neural networks. The DNN is
to train such that the OOK symbol transmitted under any given dimming target is successfully recovered over the optical channel. To generate a binary OOK symbol from the DNN, a novel stochastic binarization technique that obtains a discrete output signal with very small computational efforts is applied at the encoding neural network. Note that state-of-the-art DL libraries are typically intended for unconstrained machine learning applications. Furthermore, the training algorithm in \cite{HLee:18b,HLee:19} applies only for a single dimming target with the exact constant weight constraint. Therefore, it is not straightforward to train the proposed DNN transceiver with multiple dimming constraints by existing DNN training strategies.

For an efficient training strategy of the binary DNN transceiver, the SER minimization problem is recast into a dual optimization formulation. Dual variables associated with individual dimming constraints are optimized along with the DNN parameters via a single training step without gradient vanishing issues. This change contrasts computational efforts in the training strategy with previous training techniques where computationally demanding multiple repetitions of the training step run for the recovery of a binary output symbol. %\textcolor{red}\{The proposed DNN transceiver is extended to handle nontrivial implementation issues in the VLC systems such as nonlinear nature of LEDs and randomness of optical channel matrices.}
Numerical results demonstrate the feasibility and viability of the DNN-based design approach to the optical transceiver with arbitrary dimming support. The DNN-based design approach is verified to achieve theoretically known the best configuration of CWC codebooks under the maximum-likelihood (ML) decoding. In addition, the robustness of the DNN to some implementation issues including nonlinear properties of LEDs and time-varying optical channels is demonstrated. The proposed approach sheds a light on the design of the transceiver pair under the constrained signal constellation.

The organization of the paper is as follows: In Sec. \ref{sec:sec3}, the system model is described for the dimmable VLC systems with the OOK modulation. In addition, the design task of universal codebook is formulated in DNN's perspectives and brief preliminaries of the DNN is introduced. Sec. \ref{sec:sec4} develops a DNN structure for the VLC transceiver design, and Sec. \ref{sec:sec5} presents the corresponding constrained training strategy. Sec. \ref{sec:sec51} addresses the consideration practical issues about LED nonlinearity and time-varying optical channel effects. Sec. \ref{sec:sec6} provides numerical results that validate the efficacy of the proposed system. Sec. \ref{sec:sec7} concludes the paper.

\textit{Notations}: Throughout this paper, we employ uppercase boldface letters, lowercase boldface
letters, and normal letters for matrices, column vectors, and scalar quantities, respectively. A set of all
real matrices of size $U$-by-$V$ is defined as $\mathbb{R}^{U\times V}$. Also, $[\mathbf{z}]_{k}$ stands for the $k$-th element of a vector $\mathbf{z}$. In addition, $\mathbf{I}_{U}$ accounts for an identity matrix of size $U$-by-$U$, and $\mathbf{0}_{U}$ and $\mathbf{1}_{U}$ indicate all zero and all one column vectors of length $U$, respectively.

\section{System Model}\label{sec:sec3}

\begin{figure}
\begin{center}
\includegraphics[width=0.9\linewidth]{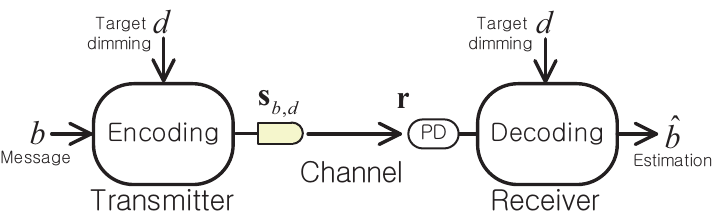}
\end{center}
%\vspace{-5mm}
\caption{Schematic diagram for OOK-modulated dimmable VLC system.}
%\vspace{-5mm}
\label{fig:fig1}
\end{figure}

Consider an OOK-modulated binary VLC system as depicted in Fig. \ref{fig:fig1}. The transmitter equipped with a white LED transmits a message $b\in\mathcal{M}\triangleq\{1,\cdots,M\}$ to the receiver with a photodetector (PD) through OOK optical pulses emitted from the transmit LED. In the OOK systems, each message $b$ is encoded to a binary vector $\mathbf{s}_{b,d}$ of length $N$ which satisfies a dimming target $d\in\mathcal{D}\triangleq\{d_{1},\cdots,d_{D}:0 \leq d_{1}<\cdots< d_{D}\leq N\}$ stemmed externally from users. The received intensity vector $\mathbf{r}\in\mathbb{R}^{N\times 1}$ at the receiver is given by
\begin{align}
    \mathbf{r}=\mathbf{H}\mathbf{s}_{b,d}+\mathbf{n},\label{eq:y}
\end{align}
where $\mathbf{H}\in\mathbb{R}^{N\times N}$ is an optical communication channel matrix and $\mathbf{n}$ stands for an additive noise vector with individual elements of zero mean and variance $\sigma^2$. The receiver obtains estimated message $\hat{b}\in\mathcal{M}$ based on the received signal vector $\mathbf{r}$. The target dimming $d$ is also known to the receiver. %for the estimation procedure such as codebook based the ML decoding.
The channel model in (\ref{eq:y}) can be generalized by including line-of-sight (LoS) and multipath components of the optimal signal. For instance, if $\mathbf{H}=\mathbf{I}_{N}$, the channel is an additive Gaussian noise channel representing the LoS environment \cite{SZhao:17}. Also, setting $\mathbf{H}$ as a Toeplitz matrix can model the inter-symbol interference (ISI) induced by the reflection of obstacles \cite{HWang:18}. %Note that indoor VLC systems with fixed LED and PD configurations undergoes the optical channel that rarely changes \cite{SZhao:17,HWang:18} so that it can be efficiently estimated via standard channel estimation techniques \cite{RWang:17}.

For the OOK dimming support, the number of nonzero symbols (or Hamming weight) of codeword $\mathbf{s}_{b,d}$ in a codebook $\mathcal{S}_{d}\triangleq\{\mathbf{s}_{b,d},\forall b\in\mathcal{M}\}$ is determined such that the average transmit intensity fulfils the dimming target $d$. There are two different approaches for the dimming control methods: strict and relaxed dimming control. The strict dimming constraint can be expressed as
\begin{align}
    \sum_{i=1}^{N}[\mathbf{s}_{b,d}]_{i}=d,\ \forall \mathbf{s}_{b,d}\in\mathcal{S}_{d},d\in\mathcal{D}. \label{eq:str_dim}
\end{align}
Here, the Hamming weight of any codeword should be same as the dimming target $d$. A codebook where all component codewords satisfy (\ref{eq:str_dim}) is known as a CWC, and mathematical properties of its family, such as the minimum Hamming distance and the dimension of the codebook, have been intensively investigated \cite{Ostergard:10}. However, the exact constant weight constraint in (\ref{eq:str_dim}) limits the number of codewords, which results in the degraded spectral efficiency and decoding error performance \cite{SHLee:15,SZhao:17}. Furthermore, the CWC is feasible only for a discrete set of dimming target values, i.e., the feasible dimming set $\mathcal{D}$ becomes $\mathcal{D}=\{1,\cdots,N\}$, and thus it fails to support arbitrary dimming requirement in practice.

To tackle these issues, the relaxed dimming constraint, which can be viewed as a generalization of the dimming constraint, has been recently addressed in \cite{SZhao:17} such that the transmit intensity averaged over codebook $\mathcal{S}_{d}$ is controlled. This constraint can be expressed as
\begin{align}
    \mathbb{E}_{\mathcal{S}_{d}}\left[\sum_{i=1}^{N}[\mathbf{s}_{b,d}]_{i}\right]=\frac{1}{M}\sum_{\mathbf{s}_{b,d}\in\mathcal{S}_{d}}\sum_{i=1}^{N}[\mathbf{s}_{b,d}]_{i}=d,\ \forall d\in\mathcal{D},\label{eq:dimming}
\end{align}
where $\mathbb{E}_{X}[\cdot]$ accounts for the average operation over a random variable $X$. We assume equally probable messages, i.e., $\Pr\{b\}=\frac{1}{M}$, $\forall b\in\mathcal{M}$. We refer to a codebook with the dimming constraint (\ref{eq:dimming}) as a semi-CWC. Since a feasible set generated from (\ref{eq:dimming}) has, in general, a larger population than a feasible set obtained from the strict dimming constraint (\ref{eq:str_dim}), as compared to the exact CWC method in \cite{Ostergard:10}, a higher degree of freedom can be expected for the semi-CWC design. It has been shown in \cite{SZhao:17} that the code rate of semi-CWCs can be enhanced over the CWC. In general, the new dimming constraint allows a higher dimensional codebook than the exact CWC (\ref{eq:str_dim}). For example, if $N=8$, $M=4$, and $d=2$, there are approximately $1.7\times 10^8$ binary codebook candidates for the semi-CWC, whereas the number of all possible exact CWC is $2\times 10^4$, which amounts only to $0.01\ \%$ with respect to the semi-CWC. Therefore, the development of a binary VLC system with the dimming constraint (\ref{eq:dimming}) turns out to be more complicated than the strict dimming constraint case.

\subsection{Formulation}
We desire to identify an efficient OOK transceiver that minimizes the average symbol error rate (SER) over the noisy channel \eqref{eq:y} for a given channel matrix $\mathbf{H}$ under the dimming constraint (\ref{eq:dimming}) and adapts universally to all possible dimming values in $\mathcal{D}$. Let $\mathbf{s}_{b,d}=f_{\text{E}}(b,d)$, $\forall b\in\mathcal{M}, \forall d\in\mathcal{D}$ and $\hat{b}=f_{\text{D}}(\mathbf{r},d)$ denote encoding and decoding functions associated with modulation and demodulation operations of the VLC transceiver, respectively. The corresponding SER minimization problem for a fixed optical channel $\mathbf{H}$ can be formulated in an optimization given by\footnote{A generalization of \eqref{eq:P1} to random channels is addressed in Sec. \ref{sec:sec5}.}
\begin{align}
    &~~\min\limits_{f_{\text{E}}(\cdot),f_{\text{D}}(\cdot)} P_{e}(\mathbf{H})\triangleq\mathbb{E}_{\mathbf{n},d,b}[\Pr\{\hat{b}\neq b|\mathbf{H}\}]\label{eq:P1}\\
    &\text{subject to}\ \frac{1}{M}\sum_{\mathbf{s}_{b,d}\in\mathcal{S}_{d}}\sum_{i=1}^{N}[\mathbf{s}_{b,d}]_{i}=d,\ \forall d\in\mathcal{D},\nonumber\\
    &~~~~~~~~~~~~ \mathbf{s}_{b,d}\in\{0,1\}^{N},\ \forall b\in\mathcal{M},\ \forall d\in\mathcal{D}.\nonumber
\end{align}
The solution of nonconvex problem (\ref{eq:P1}) is, in general, not straightforwardly obtained since no closed-form expression is available for the objective function with generic encoding and decoding rules $f_{\text{E}}(b,d)$ and $f_{\text{D}}(\mathbf{r},d)$. This poses the difficulty for solving \eqref{eq:P1} via traditional combinatorial optimization techniques, such as a branch-and-bound algorithm \cite{Boyd:04}. A brute-force approach for the identification of the optimal encoding rule $f_{\text{E}}(b,d)$ requires computationally demanding search over all possible dimming targets in $\mathcal{D}$, and the dimension of its search space is as large as $D {2^{N}\choose M}$. Furthermore, the semi-CWC design \cite{SZhao:17} that aims to maximize the code rate of $\mathcal{S}_{d}$ for a specific dimming target $d$ cannot be straightforwardly employed in solving the average SER minimization problem. Thus, random generation of CWCs has been typically utilized for the dimming control \cite{SZhao:16,SZhao:17} and obviously results in the inefficiency of the decoding performance.

\subsection{Unsupervised Learning Approach}
To handle a numerical optimization problem, a machine learning approach can be applied. In combinatorial design problem of \eqref{eq:P1}, there exist inherently multiple candidates satisfying the dimming constraint, i.e., the best mapping rule of the message index $b\in\mathcal{M}$ to the corresponding dimmable codeword $\mathbf{s}_{b,d}$ is not necessarily unique. In such a nontrivial task, a supervised learning technique does not suit very well since it mostly ends up choosing one tangible one-to-one mapping rule from an input feature to a known label, whereas the formulation in \eqref{eq:P1} identifies an unknown rule that maps one of many relationships among $b$, $\mathbf{s}_{b,d}$, and its decoded message~$\hat{b}$.

To address this challenge, we develop an unsupervised learning approach for determining an efficient semi-CWC based OOK system with universal dimming support. In particular, we obtain a DL technique that replaces unknown modulation and demodulation functions $f_{\text{E}}(b,d)$ and $f_{\text{D}}(\mathbf{r},d)$ with two DNNs: encoding and decoding neural networks, respectively. The joint design of encoding and decoding networks is referred to as an AE \cite{YBengio:09}, aiming at reconstructing the input of the encoding network from the output of the decoding network. The AE framework has recently been developed for the OOK-modulated VLC systems \cite{HLee:18b,HLee:19}. However, it is difficult to handle the DNN in \cite{HLee:18b} and \cite{HLee:19} for the universal dimming support since its neural network structure is configured for the strict dimming constraint (\ref{eq:str_dim}). In addition, the training of the DNN in \cite{HLee:18b} and \cite{HLee:19} involves a multistage annealing technique that repeats gradual increases of hyper-parameters that allow the resulting neural network to handle discrete message sets associated with a single dimming target. Although this multistage adaptation technique for hyper-parameters proves very successful in finding the optical signal set for corresponding discrete messages, it is sensitive to the hyper-parameter update schedule. Thus, the task of tuning the parameter update to training a single DNN to assimilate multiple DNNs respectively associated with all dimming targets in $\mathcal{D}$ becomes readily intractable. On the contrary, the new formulation in (\ref{eq:P1}) requires to accommodate multiple dimming constraints at the same time in a single training process. Therefore, a new strategy for the DNN construction and training process is essential for handling the design problem in (\ref{eq:P1}).

\subsection{Deep Neural Network Preliminaries} %\label{sec:sec2}
%\begin{figure}
%\begin{center}
%\includegraphics[width=0.45\linewidth]{fig0}
%\end{center}
%\caption{Deep neural network with $L$ hidden layers.}
%\label{fig:fig0}
%\end{figure}

In this section, we briefly introduce the basics of DNNs. A fully-connected DNN comprised of an input layer, multiple hidden layers, and an output layer. For a training sample $\mathbf{x}^{(j)}\in\mathbb{R}^{P\times1}$ from the training set $\mathcal{X}\triangleq\{\mathbf{x}^{(j)}:j=1,\cdots,J\}$, the DNN computes the corresponding output $\mathbf{y}^{(j)}\in\mathbb{R}^{Q\times1}$ via the cascaded layer of neural networks. %\textcolor{red}{Hence, the DNN can be represented as a mapping $\mathbf{y}=g(\mathbf{x};\Theta)$ from the input $\mathbf{x}$ to the output $\mathbf{y}$ through its parameter $\Theta$ which will be explained shortly.}
Let $L$ be the number of hidden layers. Each hidden layer $l$ for $l=1,\cdots,L$ calculates the corresponding output vector $\mathbf{h}_{l}\in\mathbb{R}^{C_{l}\times 1}$ as
\begin{align}
    \mathbf{h}_{l}=a_{l}(\mathbf{W}_{l}\mathbf{h}_{l-1}+\mathbf{b}_{l}),\ \forall l=1,\cdots,L,\label{eq:hidden}
\end{align}
where $\mathbf{h}_{0}=\mathbf{x}\in\mathcal{X}$ denotes the input of the DNN, $\mathbf{W}_{l}\in\mathbb{R}^{C_{l}\times C_{l-1}}$ and $\mathbf{b}_{l}\in\mathbb{R}^{C_{l}\times 1}$ are a weight matrix and a bias vector at hidden layer $l$, respectively, and an element-wise real-valued function $a_{l}(\cdot)$ indicates an activation that introduces nonlinear features to the output of each layer. The output layer produces the DNN output $\mathbf{y}$ from the neural network layer in (\ref{eq:hidden}) associated with an activation $a_{L+1}(\cdot)$, a weight matrix $\mathbf{W}_{L+1}\in\mathbb{R}^{Q\times C_{L}}$, and a bias vector $\mathbf{b}_{L+1}\in\mathbb{R}^{Q\times 1}$. The activation function $a_{l}(\cdot)$, the output dimension $C_{l}$, and the number of hidden layers $L$ are design parameters chosen according to applications and types of the data handled by the DNN. %\textcolor{red}{Consequently, the DNN mapping $\mathbf{y}=g(\mathbf{x};\Theta)$ can be specified by a composite function of the successive computations in \eqref{eq:hidden}.}

The DNN parameter $\Theta\triangleq\{\mathbf{W}_{l},\mathbf{b}_{l}\}$ is optimized to minimize the cost function defined for the DNN $\mathcal{C}(\Theta)$, which can be represented as
\begin{align}
    \min\limits_{\Theta} \mathcal{C}(\Theta)\triangleq\frac{1}{J}\sum_{j=1}^{J}c(\mathbf{x}^{(j)},\mathbf{y}^{(j)}). \label{eq:P0}
\end{align}
Here, a pairwise cost function $c(\mathbf{x}^{(j)},\mathbf{y}^{(j)})$  is evaluated over each training input-output pair $(\mathbf{x}^{(j)},\mathbf{y}^{(j)})$ and characterizes the purpose, e.g., between classification and regression~\cite{LeCun:15}.

The consecutive nonlinear output neural network computations and the choice of a possibly nonconvex activation in (\ref{eq:hidden}) do not allow a closed-form solution for the optimization problem in (\ref{eq:P0}). State-of-the-art DL techniques use a stochastic gradient descent (SGD) algorithm where the parameter $\Theta$ is iteratively updated in a gradient descent manner as
\begin{align}
\Theta^{[t]}=\Theta^{[t-1]}-\eta\nabla_{\Theta}\mathcal{C}(\Theta^{[t-1]}),\label{eq:SGD}
\end{align}
where $\Theta^{[t]}$ denotes the DNN parameter updated by the SGD algorithm at the $t$-th iteration, $\eta>0$ stands for a learning rate, and $\nabla_{\Theta}\mathcal{C}(\Theta)=\frac{1}{J}\sum_{\mathbf{x}^{(j)}\in\mathcal{X}}\nabla_{\Theta} c(\mathbf{x}^{(j)},\mathbf{y}^{(j)})$ indicates the gradient of the cost function with respect to the parameter $\Theta$. The gradient with respect to each weight matrix and bias vector can be computed from the back propagation (BP) algorithm based on the chain rule \cite{LeCun:15}. As a result, the gradient of the cost function with respect to $\mathbf{W}_{l}$ and $\mathbf{b}_{l}$ at hidden layer $l$ is given by the multiplication of those of the subsequent layers up to $L+1$.

Meanwhile, the calculation of the gradient $\nabla_{\Theta}\mathcal{C}(\Theta)$ would require expensive computations for a large number of training samples. For a more efficient implementation, $\nabla_{\Theta}\mathcal{C}(\Theta)$ in (\ref{eq:SGD}) is usually replaced with its approximation given by
\begin{align}
    \nabla_{\Theta}\mathcal{C}(\Theta)\simeq\frac{1}{|\mathcal{B}|}\sum_{\mathbf{x}^{(j)}\in\mathcal{B}}\nabla_{\Theta} c(\mathbf{x}^{(j)},\mathbf{y}^{(j)}),\label{eq:minibatch}
\end{align}
which is evaluated over a subset $\mathcal{B}\subset\mathcal{X}$ called a \textit{mini-batch} set. With the aid of commercial graphics processing units, the DNN parameter $\Theta$ can be efficiently determined via the SGD update in (\ref{eq:SGD}) for all possible mini-batch sets until the cost function converges to a fixed point. Once the DNN is trained, the performance is examined by calculating the cost function $\mathcal{C}(\Theta)$ with a set of test samples that have not been contained in the training set.

\section{DNN-Based Binary VLC Transceiver}\label{sec:sec4}
\begin{figure*}
\begin{center}
\includegraphics[width=.8\linewidth]{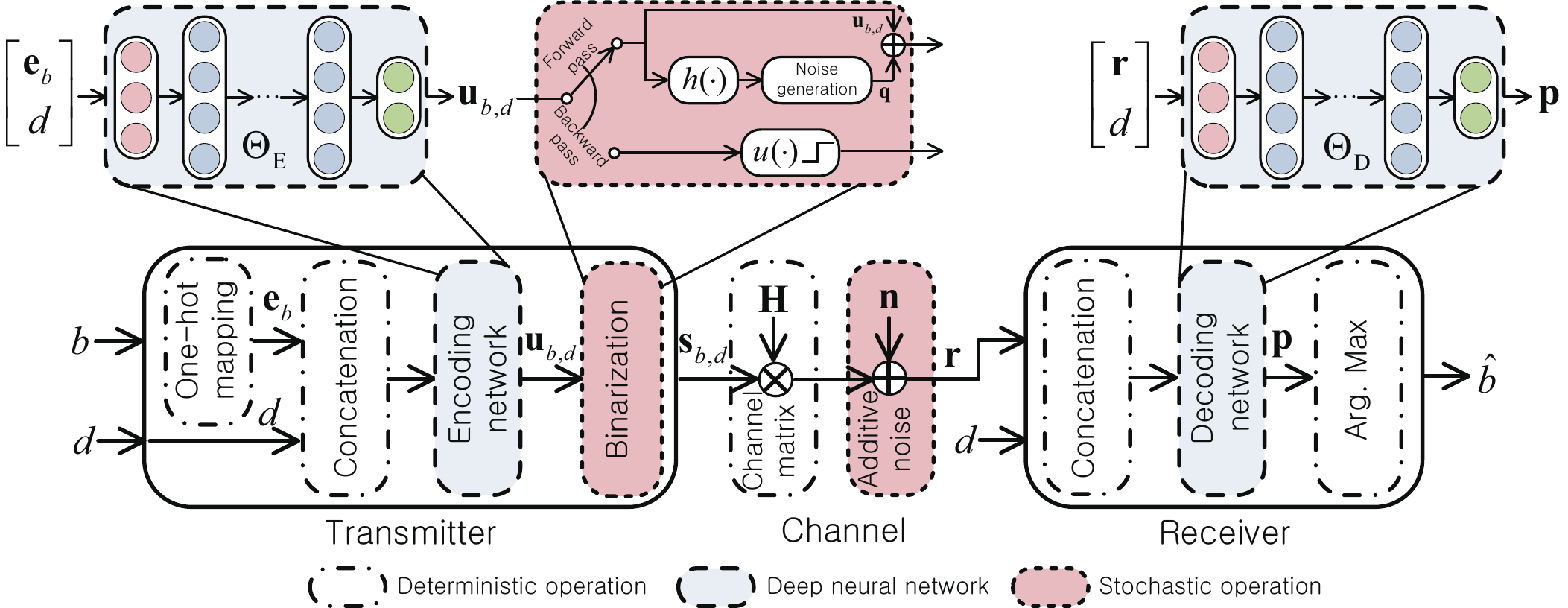}
\end{center}
%\vspace{-5mm}
\caption{Proposed DNN based dimmable OOK-modulated VLC system.}
%\vspace{-5mm}
\label{fig:fig2}
\end{figure*}

We propose a DL framework for the OOK-modulated VLC system shown in Fig. \ref{fig:fig2}. The transmitter-receiver pair is implemented with an encoding neural network, a binarization operation, an optical channel layer and a decoding neural network. The transmitter accepts an ordered pair of message and dimming $(b,d)$ for $b\in\mathcal{M}$ and $d\in\mathcal{D}$ and generates the codeword $\mathbf{s}_{b,d}$ through the binary-output encoding neural network, while the decoding neural network at the receiver side produces the estimated message $\hat{b}\in\mathcal{M}$ from the optical channel layer output $\mathbf{r}$ and the target dimming $d$. Such a DNN structure can be regarded as an AE that reconstruct the input from its corrupted channel output via subsequential encoding and decoding rules realized by DNNs. We discuss the detailed structure of the proposed DNN based transceiver.

\subsection{Encoding Neural Network}
At the transmitter, the message $b$ is first mapped to a one-hot vector representation $\mathbf{e}_{b}\in\mathbb{R}^{M\times 1}$ where the $b$-th element is equal to one and the other elements are set to zero. The one-hot representation has been widely utilized in DL classification applications that characterize class labels by multi-dimensional vectors for the neural network computation~\cite{OShea:17}.

The stacked vector $[\mathbf{e}_{b}^{T}, d]^{T}\in\mathbb{R}^{(M+1)\times 1}$ is then passed to an encoding neural network. The neural network consists of $L_{\text{E}}$ hidden layers with parameter $\Theta_{\text{E}}\triangleq\{\mathbf{W}_{\text{E},l},\mathbf{b}_{\text{E},l}\}$, where $\mathbf{W}_{\text{E},l}\in\mathbb{R}^{C_{\text{E},l-1}\times C_{\text{E},l}}$ and $\mathbf{b}_{\text{E},l}\in\mathbb{R}^{C_{\text{E},l}\times 1}$ denote the weight and bias of the encoding neural network, respectively, and $C_{\text{E},l}$ stands for the output dimension. A rectified linear unit (ReLU), a popular nonlinear activation in the DL \cite{LeCun:15}, is applied at hidden layers as
\begin{align}
    \mathbf{h}_{\text{E},l}=\max\{\mathbf{W}_{\text{E},l}\mathbf{h}_{\text{E},l-1}+\mathbf{b}_{\text{E},l},\mathbf{0}_{C_{\text{E},l}}\},\ \forall l=1,\cdots,L_{\text{E}},\nonumber
\end{align}
where $\mathbf{h}_{\text{E},l}\in\mathbb{R}^{C_{\text{E},l}\times 1}$ represents the output at hidden layer $l$ of the encoding neural network. The output dimension $C_{\text{E},l}$ of the $l$-th hidden layer for $l=1,\cdots,L_{\text{E}}$ is a hyper-parameter to be optimized through validation. For the output layer, a linear activation with the dimension $N$ is employed to produce the transmit intensity vector of length $N$. Thus, the output vector $\mathbf{u}_{b,d}$ of the encoding neural network for an input $(b,d)$ is constructed by $\mathbf{u}_{b,d}=\mathbf{W}_{\text{E},L_{\text{E}}+1}\mathbf{h}_{\text{E},L_{\text{E}}}+\mathbf{b}_{\text{E},L_{\text{E}}+1}$, which is still a continuous-valued vector.

\subsection{Binarization}\label{sec:sec4B}
To produce a binary output vector $\mathbf{s}_{b,d}$ that is mapped to an OOK-modulated optical signal, additional processing is necessary for the network output $\mathbf{u}_{b,d}$. It would be a simple binarization function, e.g., a unit-step function, used as the activation at the output layer of the encoding neural network. %such that each element $[\mathbf{s}_{b,d}]_{i}$ is equal to one if $[\mathbf{u}_{b,d}]_{i}>0$ and zero otherwise.
Such a straightforward hard thresholding technique, however, ends up with the vanishing gradient problem which is significantly challenging in training the DNNs. Specificcally, the gradient of the unit-step function is trivially zero for all input regime, and the parameter $\Theta_{\text{E}}$ for the encoding rule would not be learned from the SGD algorithm (\ref{eq:SGD}).

A recent work \cite{HLee:18b,HLee:19} uses a parameterized sigmoid activation $\text{sig}(z,\delta,0)\triangleq\frac{1}{1+\exp(-\delta (z-\Delta_d))}|_{\Delta_d=0}$ $=1/(1+\exp(-\delta z))$ to the output layer of the encoding neural network, where a nonnegative parameter $\delta$ adjusts the slope of the sigmoid function. Since the parameterized sigmoid function has nonzero gradient for a finite $\delta$, the gradient vanishing issue can be handled in training the DNN with a moderate $\delta$. From the fact that $\text{sig}(z,\delta,0)$ approximates the unit-step function accurately if $\delta$ becomes sufficiently large, the parameter $\delta$ is gradually raised during the training step to obtain the binary output \cite{HLee:18b,HLee:19}. However, it requires a very large number of iterations for cascaded training steps of the DNN with different $\delta$, incurring a slow convergence performance. Also, since the methods in \cite{HLee:18b,HLee:19} are developed for the strict dimming constraint (\ref{eq:str_dim}) through the deterministic binarization, it would not be valid for the dimming constraint (\ref{eq:dimming}) which requires a stochastic dimming control in the codebook~$\mathcal{S}_{d}$.

To overcome this difficulty, we propose a novel binarization technique that adapts to universal dimming targets from stochastic binarization \cite{Raiko:15}. In this approach, a quantization noise $\mathbf{q}$ is added to the encoding neural network output $\mathbf{u}_{b,d}$ to guarantee the binary output $\mathbf{s}_{b,d}$ as
\begin{align}
    \mathbf{s}_{b,d}=\mathbf{u}_{b,d}+\mathbf{q},\label{eq:sbd}
\end{align}
where each element $[\mathbf{q}]_{i}$ for $i=1,\cdots,N$ is chosen from the Bernoulli distribution as
\begin{align}\label{eq:q}
    [\mathbf{q}]_{i}=\begin{cases}
    1-[\mathbf{u}_{b,d}]_{i}, & \text{with probability}\ h([\mathbf{u}_{b,d}]_{i}),\\
    -[\mathbf{u}_{b,d}]_{i}, & \text{with probability}\ 1-h([\mathbf{u}_{b,d}]_{i}),
  \end{cases}
\end{align}
%\begin{align}\label{eq:q}
%    [\mathbf{q}]_{i}=\begin{cases}
%    1-[\mathbf{u}_{b,d}]_{i}, & \text{with probability}\ \text{sig}([\mathbf{u}_{b,d}]_{i}),\\
%    -[\mathbf{u}_{b,d}]_{i}, & \text{with probability}\ 1-\text{sig}([\mathbf{u}_{b,d}]_{i}),
%  \end{cases}
%\end{align}
where $h(z)\in[0,1]$ is a nonnegative function such that its integral over the range of the encoding neural network output, respectively denoted by $[\mathbf{u}_{b,d}]_{\min}$ and $[\mathbf{u}_{b,d}]_{\max}$, is equal to the ratio of the dimming target to the coderword length, i.e., $\int_{[\mathbf{u}_{b,d}]_{\min}}^{[\mathbf{u}_{b,d}]_{\max}} h(z) dz =\frac{d}{N}$. By choosing such a function $h(z)$ as the probability of the Bernoulli random variable $[\mathbf{q}]_{i}$, the dimming value for the resulting binary codeword $\mathbf{s}_{b,d}$ is approximated to $\mathbb{E}_{\mathbf{q},\mathbf{u}_{b,d}}[\sum_{i=1}^{N}[\mathbf{s}_{b,d}]_{i}]\approx N\int_0^{[\mathbf{u}_{b,d}]_{\max}} h(z) dz=d$. A possible candidate for $h(z)$ is a parameterized sigmoid function $\text{sig}(z,1,\Delta_d)=1/(1+\exp(-(z-\Delta_d)))$ with $\Delta_d \in \mathbb{R}$ chosen such that $\int_{[\mathbf{u}_{b,d}]_{\min}}^{[\mathbf{u}_{b,d}]_{\max}} \text{sig}(z,1,\Delta_d) dz=\frac{d}{N}$. We employ the sigmoid function for the stochastic binarization since it has been shown to be suitable for characterizing the binary classification probability in logistic regression problems \cite{LeCun:15}. It is obvious that each element $[\mathbf{s}_{b,d}]_{i}$ of the transmit intensity vector $\mathbf{s}_{b,d}$  in (\ref{eq:sbd}) becomes a binary number subject to stochastic noise $[\mathbf{q}]_{i}$ in (\ref{eq:q}).

The remaining task for the stochastic binarization is to identify the gradient $\nabla\mathbf{s}_{b,d}$ for training the encoding neural network parameter $\Theta_{\text{E}}$ through the SGD algorithm in (\ref{eq:SGD}). However, no closed-form expression for the gradient $\nabla\mathbf{s}_{b,d}$ exists due to the stochastic operation (\ref{eq:sbd}). To handle this, we employ an unbiased estimation on the gradient as
\begin{align}
    \nabla\mathbf{s}_{b,d}\simeq&\nabla\mathbb{E}_{\mathbf{q}}[\mathbf{s}_{b,d}|h(\mathbf{u}_{b,d})]\nonumber\\
    =&\nabla[\mathbf{u}_{b,d}+(\mathbf{1}_{N}-\mathbf{u}_{b,d})\circ h(\mathbf{u}_{b,d})\nonumber\\
    &-\mathbf{u}_{b,d}\circ(\mathbf{1}_{N}-h(\mathbf{u}_{b,d}))]=\nabla h(\mathbf{u}_{b,d}). \label{eq:grad_est}
\end{align}
It has been shown that the unbiased gradient estimation works well with the DNNs in image processing \cite{Raiko:15,YBengio:13}. Note that the gradient estimate (\ref{eq:grad_est}) is utilized only for the BP computation during the training step, while the transmitted intensity vector in (\ref{eq:sbd}) is calculated using the deterministic unit-step function, which replaces the stochastic binarization, during the verification step and is forwarded to the decoding neural network at the receiver to calculate the cost function and the estimated message $\hat{b}$.

\subsection{Decoding Neural Network}
A fixed channel matrix $\mathbf{H}$ is applied to the transmitted codeword $\mathbf{s}_{b,d}$ for considering the VLC channel impairment, and then the optical additive Gaussian noise $\mathbf{n}$ of variance $\sigma^{2}$ is added to $\mathbf{H}\mathbf{s}_{b,d}$. Based on these operations, %\textcolor{red}{We add the optical noise $\mathbf{n}$ of variance $\sigma^2$ %\sim\mathcal{N}(0,\sigma^{2}\mathbf{I}_{N})$
%to the transmitted intensity vector $\mathbf{s}_{b,d}$ such that
the proposed DNN transceiver can learn statistical properties of the optical channel (\ref{eq:y}) by itself. In the training step, the noise vector is generated at random and independently for each input $(b,d)$. At the receiver, the concatenation $[\mathbf{r}^{T}, b]^{T}\in\mathbb{R}^{(N+1)\times 1}$ of the noisy channel output $\mathbf{r}$ and the target dimming $d$ is processed by the decoding neural network with $L_{\text{D}}$ hidden layers and the parameter $\Theta_{\text{D}}\triangleq\{\mathbf{W}_{\text{D},l},\mathbf{b}_{\text{D},l}\}$, where $\mathbf{W}_{\text{D},l}\in\mathbb{R}^{C_{\text{D},l-1}\times C_{\text{D},l}}$ and $\mathbf{b}_{\text{D},l}\in\mathbb{R}^{C_{\text{D},l}\times 1}$ indicate the weight and the bias of the decoding neural network, respectively. Similar to the encoding neural network, the activations at the hidden layers of the decoding neural network are given by the ReLU function.

To obtain the estimation $\hat{b}$ for $M$ different messages $b\in\mathcal{M}$, the output layer is implemented with the softmax activation of dimension $M$. In multi-class classification applications, the softmax activation has been widely employed to characterize the probability that the DNN input belongs to each class. In this case, we desire to obtain the probability vector $\mathbf{p}\in\mathbb{R}^{M\times 1}$ from the decoding neural network where each element $[\mathbf{p}]_{b}$ can represent the posterior probability $\Pr\{b|\mathbf{r},d\}$ that each message $b\in\mathcal{M}$ has been transmitted. The output of the decoding neural network $\mathbf{p}$ is then expressed as
\begin{align}
    [\mathbf{p}]_{b}=\frac{e^{[\mathbf{v}]_{b}}}{\sum_{m=1}^{M}e^{[\mathbf{v}]_{m}}},\forall b\in\mathcal{M},\label{eq:pm}
\end{align}
where $\mathbf{v}\triangleq\mathbf{W}_{\text{D},L_{\text{D}}+1}\mathbf{h}_{\text{D},L_{\text{D}}}+\mathbf{b}_{\text{D},L_{\text{D}}+1}$ with $\mathbf{h}_{\text{D},l}\in\mathbb{R}^{C_{\text{D},l}\times 1}$ being the output at hidden layer $l$ of the decoding network. Finally, the estimate of a single binary symbol $\hat{b}$ is obtained as $\hat{b}=\arg\max_{b\in\mathcal{M}}[\mathbf{p}]_{b}$.

\section{Proposed Training Strategy}\label{sec:sec5}
%So far, the DNN based VLC transceiver structure for the OOK systems has been provided.
In this section, we present an efficient training algorithm for the proposed DNN transceiver under the dimming constraint (\ref{eq:dimming}). The proposed training algorithm includes the optimization of the encoding network parameter $\Theta_{\text{E}}$ along with the stochastic binarization (\ref{eq:sbd}) and the decoding network parameter $\Theta_{\text{D}}$.

\subsection{Reformulation Amenable to DNN}
We construct the training set $\mathcal{X}=\{(b^{(j)},d^{(j)}):b^{(j)}\in\mathcal{M}, d^{(j)}\in\mathcal{D},j=1,\cdots,J\}$. Since the goal is to minimize the SER performance, i.e., the classification error of $M$ different transmitted messages, the cost function $\mathcal{C}(\Theta_{\text{E}},\Theta_{\text{D}})$ of the proposed DNN transceiver is set to the categorical cross-entropy between the transmitted message $b\in\mathcal{M}$ and the probability vector $\mathbf{p}$ as
\begin{align}
\mathcal{C}(\Theta_{\text{E}},\Theta_{\text{D}})=-\frac{1}{J}\sum_{j=1}^{J}\log ([\mathbf{p}]_{b^{(j)}}).\label{eq:C}
\end{align}
This corresponds to the logarithm of the estimated joint probability of the correct detection of all symbols in all codewords. We next derive an explicit formulation that examines the dimming of the proposed DNN transceiver and also complement the stochastic binarization in \eqref{eq:sbd}. The average transmit intensity is obtained as
\begin{align}
    \mathbb{E}_{b,[\mathbf{q}]_{i}}\Big[[\mathbf{s}_{b,d}]_{i}\Big]&=\frac{1}{M}\sum_{b\in\mathcal{M}}\mathbb{E}_{[\mathbf{q}]_{i}}\Big[[\mathbf{s}_{b,d}]_{i}\Big]
        =\frac{1}{M}\sum_{b\in\mathcal{M}}h([\mathbf{u}_{b,d}]_{i}).\label{eq:Esbd}
\end{align}
By combining (\ref{eq:dimming}) and (\ref{eq:Esbd}), the dimming constraint function $\mathcal{F}_{d}(\Theta_{\text{E}})$ at the DNN transmitter for $d\in\mathcal{D}$ can be expressed~as
\begin{align}
    \mathcal{F}_{d}(\Theta_{\text{E}})\triangleq\frac{1}{M}\sum_{i=1}^{N}\sum_{b\in\mathcal{M}}h([\mathbf{u}_{b,d}]_{i})=d,\ \forall d\in\mathcal{D}.\label{eq:F}
\end{align}
%\begin{figure}
%\centering
%    \subfigure[Cost function]{
%        \includegraphics[width=0.55\linewidth]{fig3a.eps}\label{fig:fig3a}
%    }
%    \subfigure[Constraint function]{
%        \includegraphics[width=0.36\linewidth]{fig3b.eps}\label{fig:fig3b}
%    }
%    \caption{Evaluation of cost and constraint functions.}
%    \label{fig:fig3}
%\end{figure}
%\textcolor{blue}{It is worthwhile to remark that, unlike the cost function (\ref{eq:C}) which should be evaluated over $\mathcal{X}$, the dimming constraint function $\mathcal{F}_{d}(\Theta_{\text{E}})$ in (\ref{eq:F}) is not necessarily evaluated with the entire training set $\mathcal{X}$ since it depends only on the encoding neural network parameter $\Theta_{\text{E}}$. In practice, $\mathcal{F}_{d}(\Theta_{\text{E}})$ for each $d\in\mathcal{D}$ can be calculated from a {\em reference set} that consists of all two-tuples of message index and dimming $d$, denoted by $\mathcal{R}_{d}\triangleq\{(1,d),\cdots,(M,d)\}$. Since reference set $\mathcal{R}_{d}$ only has $M$ elements per dimming target, an efficient evaluation of $\mathcal{F}_{d}(\Theta_{\text{E}})$ is viable with the fixed reference sets at each training iteration. Fig.~\ref{fig:fig3} summarizes the resulting cost and constraint functions.}
The optimization problem formulated for training the DNN transceiver as
\begin{align}
    &~~\min\limits_{\Theta_{\text{E}},\Theta_{\text{D}}} \mathcal{C}(\Theta_{\text{E}},\Theta_{\text{D}})\label{eq:P2}\\
    &\text{subject to}\ \mathcal{F}_{d}(\Theta_{\text{E}})=d,\ \forall d\in\mathcal{D}.\nonumber
\end{align}

\subsection{Conventional Training Methods}
The explicit constraint in \eqref{eq:P2} renders it challenging to handle the DL algorithm for the problem since the DL techniques focus typically on unconstrained learning tasks. In the existing AE-based designs, the constraints on the encoding network, such as the transmit power budget, are usually handled by normalization \cite{OShea:17} and linear projection \cite{HLee:18a}. However, those approaches are valid only with convex constraints since those operations are, in fact, convex projections. In the design task of interest involving a nonconvex combinatorial constraint, simple normalization \cite{OShea:17,HLee:18a} does not guarantee the feasibility of the binary codebook. Therefore, it invokes a new training strategy that directly handles dimming constraints via stochastic binarization \eqref{eq:q}.

One naive approach for this issue includes the regularization of the cost function with a penalty parameter \cite{HLee:18b,HLee:19,MKim:18}. The corresponding unconstrained formulation can be expressed as
\begin{align}
    &\min\limits_{\Theta_{\text{E}},\Theta_{\text{D}}} \mathcal{C}(\Theta_{\text{E}},\Theta_{\text{D}})+\mu\sum_{d\in\mathcal{D}}(\mathcal{F}_{d}(\Theta_{\text{E}})-d)^{2},\label{eq:penalty}
\end{align}
where a nonnegative penalty parameter $\mu$ controls the tradeoff between improving the performance of the DNN and enforcing the satisfaction of the dimming constraint. The training performance of the DNN relies significantly on the value of the penalty parameter. If $\mu$ is small, the trained DNN based transceiver is not likely to satisfy the dimming constraint. On the other hand, if $\mu$ is too large, the categorical cross-entropy cost would not be adequately addressed during the training. It has been shown that such a regularization technique can effectively train DNNs with a single constraint \cite{HLee:18b,HLee:19,MKim:18}. However, the proposed DNN transceiver, which involves multiple dimming constraints along with the binary constraint, would not be trained effectively with (\ref{eq:penalty}) since the simultaneous satisfaction of multiple constraints cannot be controlled by a single penalty parameter. This will be clearly shown by numerical results later.

To adjust the contribution of each dimming constraint $\mathcal{F}_{d}(\Theta_{\text{E}})=d$ to the objective function, one can apply different penalty parameter $\mu_{d}$ for each $d\in\mathcal{D}$ to (\ref{eq:penalty}). In the recent DL frameworks, the optimization of such hyper parameters normally relies on trial-and-error grid search process, which is highly difficult and time-consuming. Hence, this approach may not be feasible in practice to identify multiple penalty parameters for universal dimming support.

\subsection{Proposed Training Method}\label{sec:sec5b}
We propose an efficient training strategy for the constrained training problem (\ref{eq:P2}). We first augment a regularization term to (\ref{eq:P2}) to obtain an equivalent formulation as \cite{Boyd:10}
\begin{align}
    &~~\min\limits_{\Theta_{\text{E}},\Theta_{\text{D}}}\mathcal{C}(\Theta_{\text{E}},\Theta_{\text{D}})
        +\rho\sum_{d\in\mathcal{D}}(\mathcal{F}_{d}(\Theta_{\text{E}})-d)^{2}\label{eq:P3}\\
    &\text{subject to}\ \mathcal{F}_{d}(\Theta_{\text{E}})=d,\ \forall d\in\mathcal{D}.\label{eq:P3_const}
\end{align}
It is obvious that the regularization term $\rho\sum_{d\in\mathcal{D}}(\mathcal{F}_{d}(\Theta_{\text{E}})-d)^{2}$ with nonnegative $\rho$ does not change the optimal solution, and thus (\ref{eq:P3}) is equivalent to the original problem in (\ref{eq:P2}). The augmentation in (\ref{eq:P3}) acts as an efficient enabler for the constrained DL framework and establishes a generalized formulation of the original optimization problem (\ref{eq:P2}), as shown in various equality-constrained optimization problems \cite{Boyd:10}. Compared to the conventional penalty method (\ref{eq:penalty}), the formulation in (\ref{eq:P3}) ensures the feasibility of the solution for any given $\rho$ by reflecting the constraints (\ref{eq:P3_const}). Therefore, we can fix the penalty parameter $\rho$ as an arbitrary small number to achieve a better tradeoff between the error performance and target dimming feasibility. From intensive simulation, we set $\rho=3\times10^{-6}$ for efficient training process.
%Note that a nonnegative design parameter $\rho$ adjusts the convergence speed of the computation process in obtaining the solution \cite{Boyd:03}. From intensive simulation, we have found that $\rho=3\times10^{-6}$ is an efficient choice for boosting the training procedure of this learning task.
Although the problem in (\ref{eq:P3}) is, in general, not convex and does not satisfy the Slater's condition \cite{Boyd:04}, the weak duality holds and the Lagrange duality method can be still applied to obtain a solution of (\ref{eq:P3}) efficiently. The Lagrange function of (\ref{eq:P3}) (or the augmented Lagrange function of (\ref{eq:P2}) \cite{Boyd:10}) is written by
\begin{align}
    \mathcal{L}(\Theta_{\text{E}},\Theta_{\text{D}},\boldsymbol{\lambda})=&\mathcal{C}(\Theta_{\text{E}},\Theta_{\text{D}})
        +\sum_{d\in\mathcal{D}}\lambda_{d}(\mathcal{F}_{d}(\Theta_{\text{E}})-d)\nonumber\\
        &+\rho\sum_{d\in\mathcal{D}}(\mathcal{F}_{d}(\Theta_{\text{E}})-d)^{2},\label{eq:Lag}
\end{align}
where a nonnegative parameter $\lambda_{d}$ stands for a Lagrange multiplier associated with an individual constraint of (\ref{eq:P3}) and we define dual variable vector as $\boldsymbol{\lambda}\triangleq[\lambda_{d_{1}},\cdots,\lambda_{d_{D}}]^{T}$.
The dual function $\mathcal{G}(\boldsymbol{\lambda})$ is then defined as
\begin{align}
    \mathcal{G}(\boldsymbol{\lambda})\triangleq\min\limits_{\Theta_{\text{E}},\Theta_{\text{D}}}\mathcal{L}(\Theta_{\text{E}},\Theta_{\text{D}},\boldsymbol{\lambda}),\label{eq:primal}
\end{align}
and the dual solution $\boldsymbol{\lambda}$ can be attained by solving \cite{Boyd:04}
\begin{align}
    &\max_{\boldsymbol{\lambda}}\mathcal{G}(\boldsymbol{\lambda}).\label{eq:dual}
\end{align}
%for nonnegative dual variables.

To address primal and dual optimization problems in the DNN training task, we propose an efficient constrained training algorithm which jointly updates $\{\Theta_{\text{E}},\Theta_{\text{D}}\}$ and, in turn, estimates $\boldsymbol{\lambda}$. The underlying idea is based on the primal-dual method \cite{Boyd:03}, which has been utilized for tackling traditional constrained optimization problems, to train the encoding and the decoding neural networks with multiple dimming constraints. We derive a single gradient decent optimization algorithm for the the DNN parameters and the dual variables, which is suitable for the existing DL optimization libraries to tackle the constrained training problem in (\ref{eq:P3}). At the $t$-th iteration of the proposed training technique, the DNN parameters $\Theta_{\text{E}}^{[t]}$ and $\Theta_{\text{D}}^{[t]}$ that minimizes \eqref{eq:Lag} are computed by the steepest descent as
\begin{align}
    \Theta_{\text{E}}^{[t]}&=\Theta_{\text{E}}^{[t-1]}-\eta\mathbf{f}_{\Theta_{\text{E}}}^{[t-1]},\label{eq:pup1}\\
    \Theta_{\text{D}}^{[t]}&=\Theta_{\text{D}}^{[t-1]}-\eta\mathbf{f}_{\Theta_{\text{D}}}^{[t-1]},\label{eq:pup2}
\end{align}
where $\mathbf{f}_{\Theta_{\text{E}}}^{[t]}\triangleq\nabla_{\Theta_{\text{E}}}\mathcal{L}(\Theta_{\text{E}},\Theta_{\text{D}}^{[t]},\boldsymbol{\lambda}^{[t]})$, $\mathbf{f}_{\Theta_{\text{D}}}^{[t]}\triangleq\nabla_{\Theta_{\text{D}}}\mathcal{L}(\Theta_{\text{E}}^{[t]},\Theta_{\text{D}},\boldsymbol{\lambda}^{[t]})$, and $\boldsymbol{\lambda}^{[t]}$ stands for the dual solution at the $t$-th iteration. The gradients $\mathbf{f}_{\Theta_{\text{E}}}^{[t]}$ and $\mathbf{f}_{\Theta_{\text{D}}}^{[t]}$ are obtained using
\begin{align}
    \mathbf{f}_{\Theta_{\text{E}}}^{[t]}=&\nabla_{\Theta_{\text{E}}}\mathcal{C}(\Theta_{\text{E}},\Theta_{\text{D}}^{[t]})\nonumber\\
    &+\sum_{d\in\mathcal{D}}\Big(\lambda_{d}^{[t]}+2\rho(\mathcal{F}_{d}(\Theta_{\text{E}})-d)\Big)\nabla_{\Theta_{\text{E}}}\mathcal{F}_{d}(\Theta_{\text{E}}),\nonumber\\
    \mathbf{f}_{\Theta_{\text{D}}}^{[t]}=&\nabla_{\Theta_{\text{D}}}\mathcal{C}(\Theta_{\text{E}}^{[t]},\Theta_{\text{D}}).
\end{align}
The dual update for the dual problem (\ref{eq:dual}) is subsequently obtained as
\begin{align}
    \boldsymbol{\lambda}^{[t]}=\boldsymbol{\lambda}^{[t-1]}+\eta\mathbf{f}_{\boldsymbol{\lambda}}^{[t-1]},\label{eq:dup}
\end{align}
where the gradient $\mathbf{f}_{\boldsymbol{\lambda}}^{[t]}\triangleq\nabla_{\boldsymbol{\lambda}}\mathcal{L}(\Theta_{\text{E}}^{[t]},\Theta_{\text{D}}^{[t]},\boldsymbol{\lambda})$ is simply given by
$\mathbf{f}_{\boldsymbol{\lambda}}^{[t]}=[\mathcal{F}_{d_{1}}(\Theta_{\text{E}}^{[t]})-d_{1},\cdots,\mathcal{F}_{d_{D}}(\Theta_{\text{E}}^{[t]})-d_{D}]^{T}$. By combining \eqref{eq:pup1}, \eqref{eq:pup2}, and \eqref{eq:dup}, iterative updates can be consolidated as
\begin{align}\label{eq:up}
    \mathcal{Z}^{[t]}\triangleq\begin{bmatrix}
           \Theta_{\text{E}}^{[t]} \\
           \Theta_{\text{D}}^{[t]} \\
           \boldsymbol{\lambda}^{[t]}
         \end{bmatrix}
         = \mathcal{Z}^{[t-1]}-\eta
         \begin{bmatrix}
           \mathbf{f}_{\Theta_{\text{E}}}^{[t-1]} \\
           \mathbf{f}_{\Theta_{\text{D}}}^{[t-1]} \\
           -\mathbf{f}_{\boldsymbol{\lambda}}^{[t-1]}
         \end{bmatrix}.
\end{align}
Note that DNN parameters and dual variables can be calculated in a single SGD update rule. The update rule in (\ref{eq:up}) allows to readily employ existing mini-batch SGD algorithms, e.g., the Adam algorithm \cite{Kingma:15}, for training the DNN-based OOK transceiver. We summarize a mini-batch based training algorithm for problem (\ref{eq:P2}) in Algorithm~1. The proposed algorithm corresponds to an unsupervised learning strategy since no prior knowledge regarding the optimal encoding and decoding rules is necessary in the training step. At each training iteration $t$, the performance of the DNN transceiver is assessed by evaluating $\mathcal{L}(\Theta_{\text{E}}^{[t]},\Theta_{\text{D}}^{[t]})$ over a validation set. The validation set contains numerous realizations of message indices, dimming values, noise vectors, and channel matrices which are generated independently of the training data. Similarly with the subgradient method \cite{Boyd:03}, the validation performance is under tracking, and the DNN parameters $\Theta_{\text{E}}^{[t]}$ and $\Theta_{\text{D}}^{[t]}$ associated with the lowest Lagrangian $\mathcal{L}_{best}$ under the satisfaction of all dimming constraints are stored. This can also be viewed as an early stopping technique in the DL area \cite{Goodfellow:16} that terminates the training step if the validation performance is not improved within a few iterations. For validation and testing steps, stochastic binarization layer \eqref{eq:q} is replaced with a deterministic unit step function \cite{Raiko:15} to avoid the quantization error that occurs in the codewords utilized in the real-time operation. Then, in the testing step, learned codewords $\mathbf{s}_{b,d}$, $\forall b\in\mathcal{M},d\in\mathcal{D}$, as well as their achievable dimming levels depend only on two-tuple $(b,d)$. This guarantees the feasibility of the trained network for arbitrary test noise and channel setup.

Once the proposed DNN transceiver is trained through Algorithm~1, the encoding and decoding neural networks are split and used as the corresponding VLC transmitter and receiver, respectively. Individual neural networks can be implemented in memory units that store the DNN parameters $\Theta_{\text{E}}$ and $\Theta_{\text{D}}$ and in processors that evaluate arithmetic operations and activations. With the trained parameters at hand, the DNNs that run neural network computations (\ref{eq:hidden}) perform the modulation and the demodulation of the transmitter-receiver pair. Thus, the real-time computational complexity of the encoding and decoding networks depends on the dimension and the number of hidden layers and is given by $\mathcal{O}(MC_{\text{E},1}+\sum_{l=2}^{L_{\text{E}}}C_{\text{E},l-1}C_{\text{E},l}+C_{\text{E},L_{\text{E}}}N)$ and $\mathcal{O}(NC_{\text{D},1}+\sum_{l=2}^{L_{\text{D}}}C_{\text{D},l-1}C_{\text{D},l}+C_{\text{D},L_{\text{D}}}M)$, respectively. Since the encoding network maps a message-dimming index pair $(b,d)$ to binary codeword $\mathbf{s}_{b,d}$, its real-time operation can be simply realized as a codebook-based modulation. Hence, the complexity of the proposed DNN-based VLC transceiver depends mainly on the decoding network.

\begingroup
\renewcommand{\baselinestretch}{1.2}
\begin{algorithm}
\caption{Proposed training algorithm for (\ref{eq:P2})}
\begin{algorithmic}
    \STATE Initialize $t=0$, $\mathcal{L}_{best}=\infty$, $\Theta_{\text{E}}^{[0]}$, $\Theta_{\text{D}}^{[0]}$, and $\boldsymbol{\lambda}^{[0]}$.
    \REPEAT{}
        \FOR{each mini-batch $\mathcal{B}\subset\mathcal{X}$}
            \STATE Set $t\leftarrow t+1$.
            \STATE Update the parameter $\mathcal{Z}^{[t]}$ from (\ref{eq:minibatch}) and (\ref{eq:up}).
            \STATE Evaluate the Lagrangian $\mathcal{L}(\Theta_{\text{E}}^{[t]},\Theta_{\text{D}}^{[t]})$ over the validation data set.
            \IF{$\mathcal{L}(\Theta_{\text{E}}^{[t]},\Theta_{\text{D}}^{[t]})\leq \mathcal{L}_{best}$ and $\Theta_{\text{E}}^{[t]}$ is feasible}
                \STATE Set $\mathcal{L}_{best}=\mathcal{L}(\Theta_{\text{E}}^{[t]},\Theta_{\text{D}}^{[t]})$ and save the DNN parameters $\Theta_{\text{E}}^{[t]}$ and $\Theta_{\text{D}}^{[t]}$.
            \ENDIF
        \ENDFOR
    \UNTIL convergence
\end{algorithmic}
\end{algorithm}
\endgroup

\begin{rem}
The learned OOK codewords are regarded as dimming control codes, targeting to fulfill the dimming requirement imposed by external users. Additional error correcting capability can be provided by an outer code. Such a concatenated coding approach has been investigated in VLC systems for the multi-fold purpose of dimming support and channel coding \cite{SZhao:16,SZhao:16b}. In the DNN-based transceiver, the channel encoder and decoder can be readily attached before the encoding network and after the decoding network, respectively.
\end{rem}

\section{Consideration of LED Nonlinearity and Varying Optical Channels}\label{sec:sec51}
Previous sections have addressed the DL approach for the VLC under the assumption that the optical channel $\mathbf{H}$ is fixed and known before transmission. This would be valid in practical indoor VLC environment where the optical channel rarely changes. However, in some cases such as mobile communication, the communication channel may vary at each transmission and cannot be perfectly known to the receiver. Furthermore, material properties of practical LEDs, such as LED memory effect and nonlinear electro-to-opto transfer function \cite{XDeng:18,WZhao:18}, also have nonlinear impacts on the performance. The resulting received signal can be expressed as
\begin{align}
\mathbf{r}=\mathbf{H}g(\mathbf{s}_{b,d})+\mathbf{n},\label{eq:y_nl}
\end{align}
where the electro-to-opto transfer function $g(\mathbf{z})$ is introduced in \eqref{eq:y_nl} to characterize the LED nonlinearity. For instance, the transfer function $g(\mathbf{z})$ of a commercial LED, e.g., Kingbright blue T-1 3/4 \cite{Kbright:18}, can be represented by the Hammerstein model as \cite{EEskinat:91}
\begin{align}
[g(\mathbf{z})]_{i}=\sum_{k=1}^{K}a_{k}([\mathbf{z}]_{i})^{k}+\zeta\sum_{k=1}^{K}a_{k}([\mathbf{z}]_{i-1})^{k},\label{eq:led_nl}
\end{align}
where $K$ stands for the polynomial exponent for the nonlinearity, $a_{k}$ for $k=1,\cdots,K$ are the polynomial coefficients, and $\zeta$ denotes the LED memory factor. If $K=1$, $a_{1}=1$, and $\zeta=0$, \eqref{eq:led_nl} reduces to the identity function $g(\mathbf{z})=\mathbf{z}$, valid for an ideal linear LED without the memory effect, as in \eqref{eq:y}. The memory effect of practical LEDs makes existing dimming binary coding techniques \cite{Ostergard:10,SZhao:16,SZhao:17}, which adjust the Hamming weight of binary codeword $\mathbf{s}_{b,d}$, no longer valid since the dimming target in \eqref{eq:led_nl} is not simply determined based on the average Hamming weight but the average optical power $\mathbb{E}_{\mathcal{S}_{d}}[g(\mathbf{s}_{b,d})]$. %conveyed through the optical channels.}

To handle these practical challenges of handling time-varying and nonlinear channel environment, we extend the DNN transceiver design given in Fig. \ref{fig:fig2}. A key idea is to train the encoding and decoding networks along with nonlinear transfer function \eqref{eq:led_nl} as well as numerous channel samples capturing dynamics of optical channels. To this end, we need to find an efficient DNN structure suitable for the nonlinear LED and varying channel scenario. In the following, the detailed strategies are given for the encoding network, channel layer, and decoding network.

\subsection{Encoding network}
In IM/DD-based systems, the input current modulated by the binary electrical signal based on binary codeword $\mathbf{s}_{b,d}$ is applied to the LED \cite{XDeng:18,WZhao:18}, whereas the corresponding LED output $g(\mathbf{s}_{b,d})$ is the optical power transmitted through the optical channel \eqref{eq:y_nl}. Thus, it is still necessary to force the LED input current $\mathbf{s}_{b,d}$ to be of binary alphabets, which can be accomplished by the proposed stochastic binarization. At the transmitter, $g(\mathbf{z})$ in \eqref{eq:led_nl} is used as additional activation function so that the encoding network can learn the nonlinearity. This slight modification allows the output of the transmitter $g(\mathbf{s}_{b,d})$ to reflect the nonlinear LED optical power, and binary symbol $\mathbf{s}_{b,d}$ is utilized for designing OOK-modulated current input to the LED \cite{XDeng:18}.

The dimming support is now achieved by considering the optical power $g(\mathbf{s}_{b,d})$ in the dimming constraint \eqref{eq:dimming}, which is changed as
\begin{align}
    \frac{1}{M}\sum_{\mathbf{s}_{b,d}\in\mathcal{S}_{d}}\sum_{i=1}^{N}[g(\mathbf{s}_{b,d})]_{i}=d,\ \forall d\in\mathcal{D}.\label{eq:dim_nl}
\end{align}
The nonlinearity, in particular, the LED memory, affects the dimming value in \eqref{eq:dim_nl}, and the Hamming weight of $\mathbf{s}_{b,d}$ is no longer directly proportional to the average intensity of the LED. The SER minimization becomes challenging since existing CWC design strategies \cite{Ostergard:10,SZhao:16} with the strict dimming constraint are not feasible. The pre-distortion technique in \cite{XDeng:18} can be applied to compromise the LED nonlinearity. However, the input current to transmitter is no longer binary and violates the OOK modulation rule.

The LED nonlinearity can be captured by the DNN transceiver through the proposed end-to-end training approach. The dimming constraint function $\mathcal{F}_{d}(\Theta_{E})$ in \eqref{eq:F} is modified as
\begin{align}
    \mathcal{F}_{d}(\Theta_{\text{E}})=\frac{1}{M}\sum_{i=1}^{N}\sum_{b\in\mathcal{M}}[g(h(\mathbf{u}_{b,d}))]_{i},\quad \forall d\in\mathcal{D}.\label{eq:F_nl}
\end{align}
Since $g(\mathbf{z})$ in \eqref{eq:led_nl} is a polynomial for arbitrary $K$, $a_{k}$, and $\zeta$, the gradient of \eqref{eq:F_nl} can be readily calculated through the state-of-the-art DL libraries.

\subsection{Optical channel}
The minimization task of the SER performance averaged over the optical channel realization $\mathbf{H}$, i.e., $\mathbb{E}_{\mathbf{H}}[P_{e}(\mathbf{H})]$, is carried out to consider  nonlinear varying channels in \eqref{eq:y_nl}. Both channel matrix $\mathbf{H}$ and additive noise $\mathbf{n}$ are generated at random during the training step to evaluate a channel-averaged classification error. This is achieved by modifying the optical channel layer in Fig. \ref{fig:fig2} such that the stochastic optical channel matrix is applied to the LED power $g(\mathbf{s}_{b,d})$ obtained from each training sample. As a result, the categorical cross-entropy in \eqref{eq:C} is readily employed to quantify the channel-averaged SER $\mathbb{E}_{\mathbf{H}}[P_{e}(\mathbf{H})]$.

\subsection{Decoding network}
In VLC systems, the channel state information (CSI) is available at the receiver with the aid of the pilot-based channel estimation \cite{SVappangi:18}. Hence, the decoding network should be modified to exploit it. We consider two different strategies according to the level of the knowledge: perfect CSI and no CSI. First, in the perfect CSI case, the decoding network accepts $\mathbf{H}$ as additional input. In particular, the input to the decoding network becomes the concatenation of three-tuple $[\mathbf{r}^{T},b,\text{vec}(\mathbf{H})^{T}]^{T}\in\mathbb{R}^{(N^{2}+N+1)\times 1}$, where $\text{vec}(\cdot)$ indicates the vectorization operation. Then, it is produced by similar neural network computations in Sec. \ref{sec:sec4}. With the CSI at hand, adaptive demodulation is available for the decoding network. On the other hand, in an information-limited case where no CSI is available at the receiver, additional signaling for the channel estimation would be a burden for a practical system implemented with simple LED and PD configuration, where complex and negative transmission is not affordable. Thus, a blind detection without the CSI knowledge at the receiver is required for this case which is much challenging compared to the perfect CSI case. To this end, we train the decoding network by excluding the CSI from its input vector. Therefore, the structure of the decoding network remains identical to that of the fixed CSI setup in Sec. \ref{sec:sec4} with the exception that $\mathbf{H}$ is randomly generated for each training sample while $\mathbf{H}$ is considered as a parameter set before training in the fixed CSI scenario. The decoding network does not know the exact value of the channel matrix $\mathbf{H}$ during both training and testing steps. However, statistical properties of the dynamics of the CSI can be learned by the DNN transceiver by observing numerous received signal vectors generated from training samples. Therefore, the end-to-end VLC transceiver pair is trained such that the transmitted message can be recovered over varying optical channels without the knowledge of the CSI. Finally, the trained DNN transceiver is utilized for the blind detection of the VLC systems.

The constrained training algorithm proposed in Sec. \ref{sec:sec5} can be readily applied to train the DNN transceiver for LED nonlinearity and varying optical channels with the modified dimming constraint function \eqref{eq:F_nl}. The trained encoding network can be simply realized as a lookup table that maps message-dimming index pair $(b,d)$ to the corresponding LED input current $\mathbf{s}_{b,d}$. The decoding network can also be used for the real-time detection of the received signal in \eqref{eq:y_nl}.

\begin{rem}
There are several candidates of analytical channel models characterizing statistical propagation environment of practical VLC systems. Thus, multiple DNNs for all possible channel models are trained in advance, and, on a real-time VLC operation, one DNN is chosen with the best-matching channel statistics. Following this strategy, we can implement the DNN transceivers for different channel statistics without additional training and hyperparameter tuning.
\end{rem}

\begin{rem}
Flicker, a change in the light intensity perceived by human eye, can also be considered in the system design. To avoid it, interleaving techniques can be used for the codebook \cite{SHLee:12,JFang:17}. Once the training step is finished, the output of the encoding network is shuffled so that the low-frequency component amount of the resulting frequency spectrum is minimized \cite{JFang:17,CMejia:17}. Although the training step could be configured to obtain an interleaved codebook, it is well known that DL techniques have weak capability of enumerating or sorting the data, and significant improvement is not expected in learning results albeit incurring additional computational efforts.
\end{rem}

\section{Numerical Results}\label{sec:sec6}
In this section, we describe the structure of the proposed DNN first and test its performance by comparison of the computer simulation results with existing techniques. We construct the encoding network with $L_{\text{E}}=3$ hidden layers each of which has the dimension $2M^2$, $M^2$, and $\frac{M^{2}}{2}$, respectively, while the decoding network has a reversed structure to the encoding network such that there are $L_{\text{D}}=3$ hidden layers with the output dimension of $\frac{M^{2}}{2}$, $M^2$, and $2M^2$, respectively, unless otherwise stated.\footnote{The complexity of the DNN transceiver is given by $\mathcal{O}(M^{3}+NM^{2})$, which is comparable to the ML decoder of $\mathcal{O}(N^{2}M)$ for moderate $M$ and $N$.} The Adam algorithm \cite{Kingma:15} is adopted to train the DNNs, and the number of training and the validation samples are set to $5\times10^5\times M$. The message indices, the target dimming, and the noise vectors are randomly generated for the training, the validation, and the testing steps with different random seeds. We use the batch normalization technique \cite{Ioffe:15} at each layer of the encoding and the decoding networks for improving the training performance. The training process of the proposed DNN transceiver is implemented in Python 3.7 with TensorFlow 1.14.0.

In simulation, $N=8$ is considered\footnote{Such a short dimming control coding with $N=8$ has been typically adapted in the literature \cite{SZhao:16,SZhao:17,SZhao:16b}.} and the dimming target is given as $\mathcal{D}=\{2,2.5,3,3.5,4\}$\footnote{The OOK codebook $\mathcal{S}_{d}$ designed for a specific $d<N$ is simply modified to obtain $\mathcal{S}_{N-d}$ by flipping binary symbols in each codeword. Hence, this setup results in the encoder supporting 9 dimming levels as considered in \cite{XLiang:17,SZhao:16,SZhao:17}.}, unless otherwise stated. In this configuration, there are total $5\times10^{26}$ possible semi-CWC candidates for $M=16$, and thus a brute-force search approach for directly solving (\ref{eq:P1}) would not be practical. The signal-to-noise ratio (SNR) is defined as $\text{SNR}\triangleq\frac{E_{s}}{\sigma^{2}}=\frac{d}{N\sigma^{2}}$, where $E_{s}\triangleq\frac{\sum_{i=1}^{N}\mathbb{E}_{b}[[g(\mathbf{s}_{b,d})]_{i}]}{N}$ accounts for the symbol energy per each dimension. We train the DNN transceiver at a certain noise variance $\sigma^{2}$, which is determined via the validation, while its performance is evaluated for all SNR regimes.

\subsection{Ideal LED and Fixed Channel Setup}
\begin{figure}
\begin{center}
\includegraphics[width=.93\linewidth]{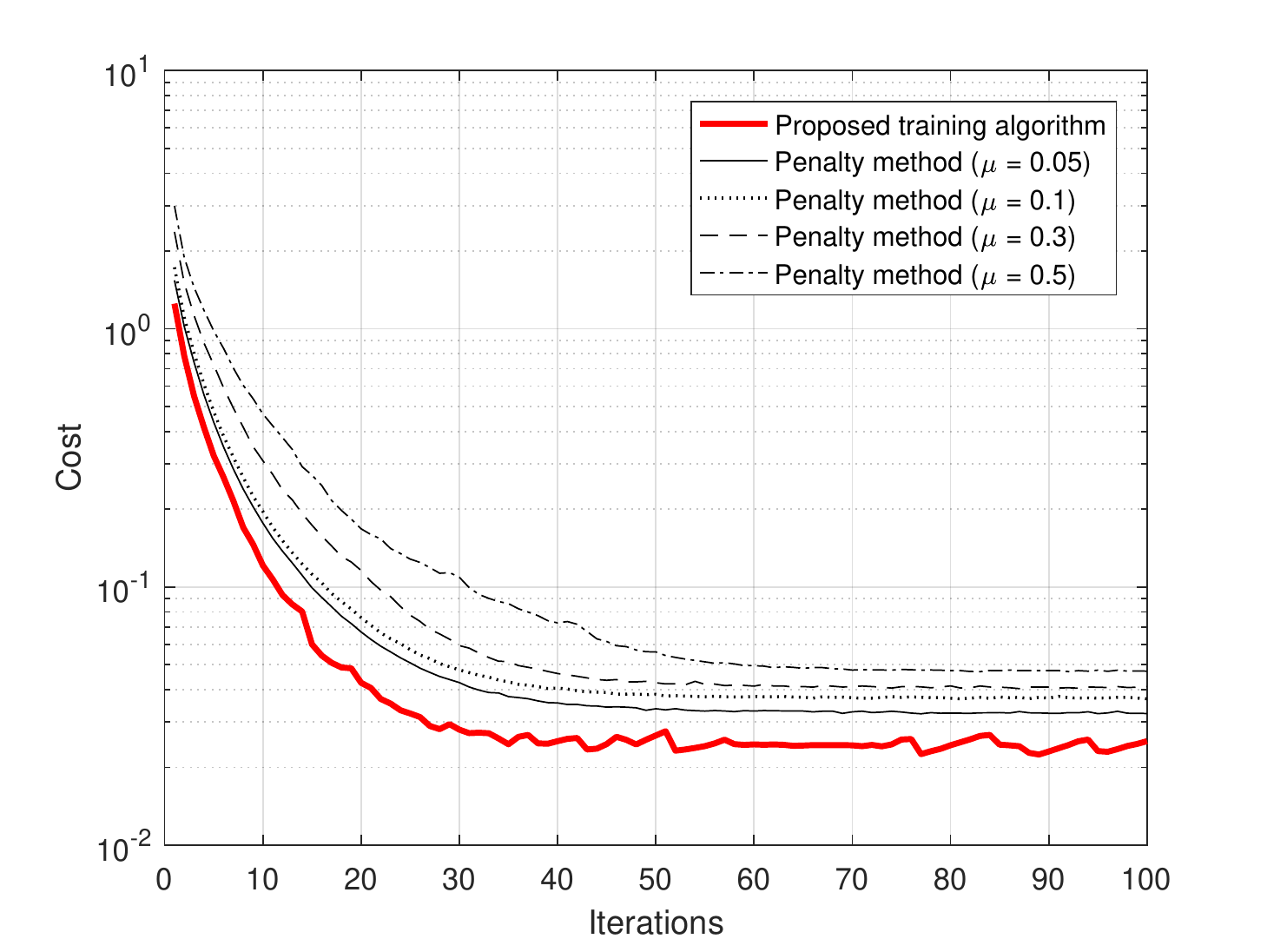}
\end{center}
%\vspace{-7mm}
\caption{Convergence behavior of the proposed training algorithm for $N=8$, $k=3$, and $\sigma^{2}=0.1$.}
%\vspace{-5mm}
\label{fig:fig4}
\end{figure}

For comparison with theoretically known best CWC configurations under ML decoding presented in \cite{Ostergard:10}, the neural network is first trained to find a codebook of good minimum distance property under the ideal linear LED and LoS setup. In other words, we consider the optical channel model in \eqref{eq:y} with fixed channel $\mathbf{H}=\mathbf{I}_{N}$ \cite{SZhao:17}, as is done for the decoding-error rate performance test in channel coding.

Fig. \ref{fig:fig4} presents the convergence of the cost function during the training step for $N=8$, $k=3$, and $\sigma^{2}=0.1$, where $k\triangleq\log_{2}M$ denotes the number of bits per symbol. For comparison, the performance of the conventional penalty method (\ref{eq:penalty}) is also plotted with different penalty parameters $\mu$. According to intensive simulations, $\mu=0.05$ is the smallest empirically found value that allows a feasible DNN for all dimming constraints, and thus the choice of $\mu=0.5$ leads to a good tradeoff between the classification performance and the dimming feasibility in the penalty method (\ref{eq:penalty}). The cost function of the proposed algorithm decays as the number of iterations grows, implying that the gradient estimation (\ref{eq:grad_est}) for stochastic binarization and the proposed training rule (\ref{eq:up}) work in practice. Also, the proposed training algorithm outperforms the penalty method regardless of $\mu$. This points out that a single penalty parameter is not sufficient for handling multiple constraints. By contrast, the proposed algorithm can effectively address multiple dimming constraints by optimizing dual variable along with the DNN parameters.

\begin{table}[]
\centering
\caption{Minimum Hamming distance performance.}\label{table:table1}
\begin{tabular}{c|c|c|c||c|c|c|c|c|}
\cline{2-9}
\multicolumn{1}{l|}{}                 & \multicolumn{3}{c||}{CWC \cite{Ostergard:10}} & \multicolumn{5}{c|}{Proposed DNN transceiver} \\ \hline
\multicolumn{1}{|c||}{$k\backslash d$} &~2~                     &~3~                     &~4~                     &~2~      & 2.5      &~3~      & 3.5      &~4~     \\ \hline\hline
\multicolumn{1}{|c||}{2}               &~4~                     &~4~                     &~4~                     &~4~      & 4        &~4~      & 4        &~\bf{5}~     \\ \hline
\multicolumn{1}{|c||}{3}               &~2~                     &~4~                     &~4~                     &~2~      & \bf{3}        &~4~      & 4        &~4~     \\ \hline
\multicolumn{1}{|c||}{4}               &~2~                     &~2~                     &~2~                     &~2~      & 2        &~2~      & \bf{3}        &~\bf{4}~     \\ \hline
\end{tabular}
\end{table}

\begin{table}[h!]
\centering
\caption{Learned Codebooks for $N=8$}
\subtable[$k=2$ and $d=4$]{
\centering
\begin{tabular}{|c|c|}
\hline
Index & Codeword                                    \\ \hline
1     & 0     1     1     1     0     1     1     1 \\ \hline
2     & 1     0     0     0     0     1     0     1 \\ \hline
3     & 0     1     0     1     1     0     0     0 \\ \hline
4     & 1     0     1     0     1     0     1     0 \\ \hline
\end{tabular}
}
\subtable[$k=3$ and $d=2.5$]{
\centering
\begin{tabular}{|c|c|}
\hline
Index & Codeword                                    \\ \hline
1     & 1     0     0     1     0     0     0     1 \\ \hline
2     & 0     1     0     1     0     0     0     0 \\ \hline
3     & 0     1     0     0     0     0     1     1 \\ \hline
4     & 0     0     0     1     1     0     1     0 \\ \hline
5     & 0     0     1     0     0     0     0     1 \\ \hline
6     & 1     0     1     0     0     0     1     0 \\ \hline
7     & 1     1     0     0     1     0     0     0 \\ \hline
8     & 0     0     0     0     0     1     0     0 \\ \hline
\end{tabular}
}
\subtable[$k=4$ and $d=3.5$]{
\centering
\begin{tabular}{|c|c|}
\hline
Index & Codeword                                    \\ \hline
1     & 1     1     0     0     1     0     0     1 \\ \hline
2     & 1     0     1     0     1     0     1     0 \\ \hline
3     & 0     0     0     1     0     1     1     0 \\ \hline
4     & 0     0     1     1     0     0     0     0 \\ \hline
5     & 0     1     0     0     0     0     1     0 \\ \hline
6     & 0     0     0     0     0     1     0     1 \\ \hline
7     & 0     0     0     1     1     0     1     1 \\ \hline
8     & 0     1     1     1     0     0     1     1 \\ \hline
9     & 0     1     1     0     1     1     1     1 \\ \hline
10     & 1     0     0     1     1     1     0     0 \\ \hline
11     & 1     0     1     0     0     1     1     1 \\ \hline
12     & 0     1     0     1     1     0     0     0 \\ \hline
13     & 1     0     0     0     0     0     0     0 \\ \hline
14     & 0     1     1     0     0     1     0     0 \\ \hline
15     & 1     1     0     1     0     1     0     1 \\ \hline
16     & 0     0     1     0     1     0     0     1 \\ \hline
\end{tabular}
}
\subtable[$k=4$ and $d=4$]{
\centering
\begin{tabular}{|c|c|}
\hline
Index & Codeword                                    \\ \hline
1     & 0     1     0     1     1     0     0     1 \\ \hline
2     & 0     0     1     0     1     0     1     1 \\ \hline
3     & 0     1     1     1     1     1     1     0 \\ \hline
4     & 0     0     1     1     0     0     0     0 \\ \hline
5     & 0     1     0     0     0     0     1     0 \\ \hline
6     & 0     0     0     1     0     1     1     1 \\ \hline
7     & 1     0     0     1     1     0     1     0 \\ \hline
8     & 1     1     1     1     0     0     1     1 \\ \hline
9     & 1     1     0     0     1     1     1     1 \\ \hline
10     & 0     0     0     0     1     1     0     0 \\ \hline
11     & 1     0     1     0     0     1     1     0 \\ \hline
12     & 1     1     1     0     1     0     0     0 \\ \hline
13     & 1     0     0     0     0     0     0     1 \\ \hline
14     & 0     1     1     0     0     1     0     1 \\ \hline
15     & 1     1     0     1     0     1     0     0 \\ \hline
16     & 1     0     1     1     1     1     0     1 \\ \hline
\end{tabular}
}
\label{table:table2}
\end{table}

Table \ref{table:table1} shows the minimum Hamming distance of the OOK codebook learned by the proposed DNN transceiver for $N=8$ and $k\in\{2,3,4\}$. For reference, theoretical maximum of the minimum Hamming distance of the CWCs \cite{Ostergard:10} are also provided for integer dimming values $d\in\{2,3,4\}$. It is first noted that, unlike the CWC developed for the strict dimming (\ref{eq:str_dim}), the proposed DNN approach generates the semi-CWC under the dimming constraint (\ref{eq:dimming}) which can achieve non-integer dimming $d=2.5$ and $3.5$. Also, in some cases highlighted by boldface letters in Table \ref{table:table1}, the DNN transceiver improves the minimum Hamming distance performance of the classical CWC design. This shows that the dimming constraint (\ref{eq:dimming}) offers an increased degree of freedom in the OOK codebook design. Several learned semi-CWCs with the enhanced minimum Hamming distance performance are listed in Table \ref{table:table2}.

\begin{figure}
\centering
    \subfigure[$k=2$]{
        \includegraphics[width=.93\linewidth]{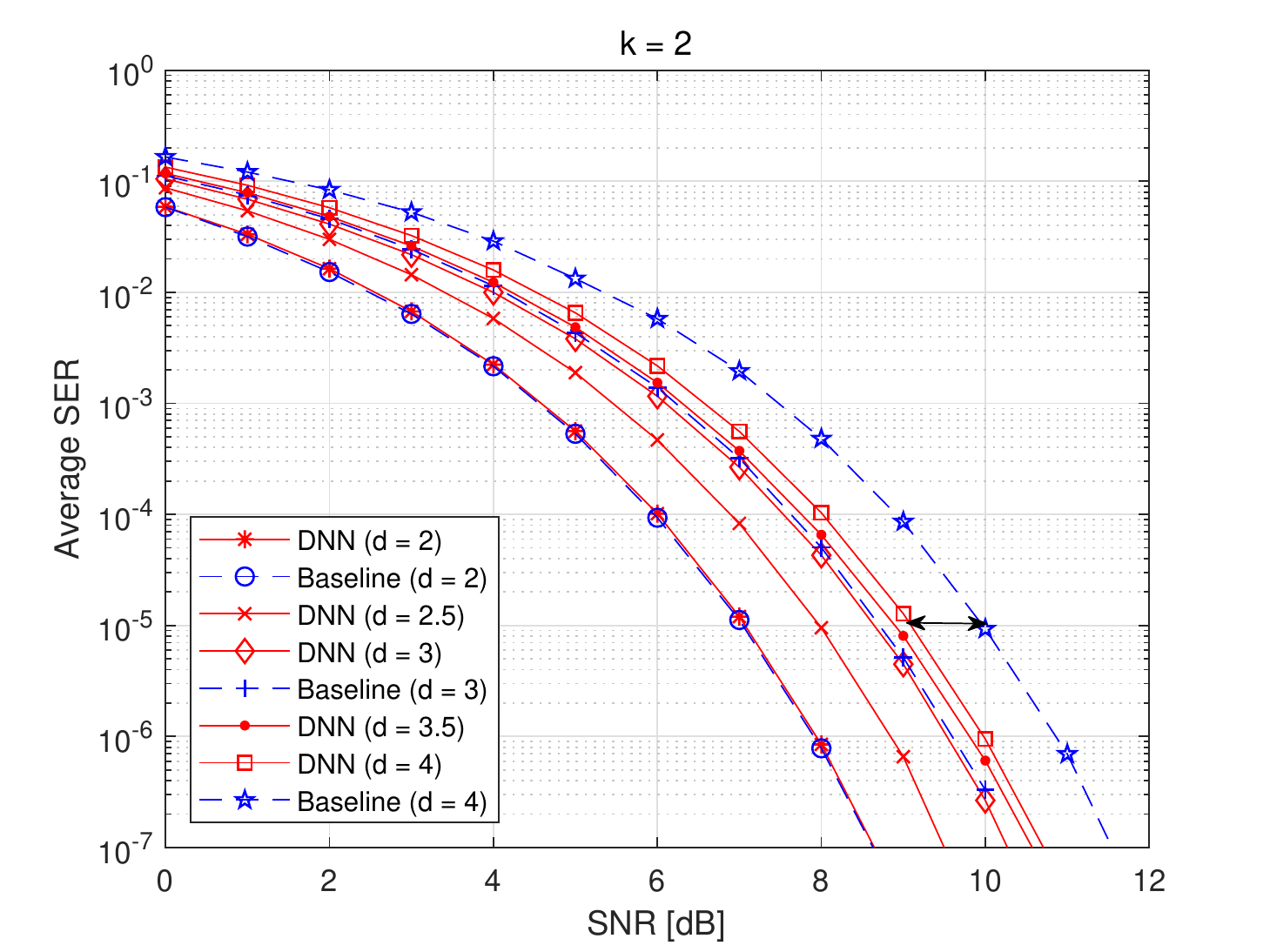}
    }
    \subfigure[$k=3$]{
        \includegraphics[width=.93\linewidth]{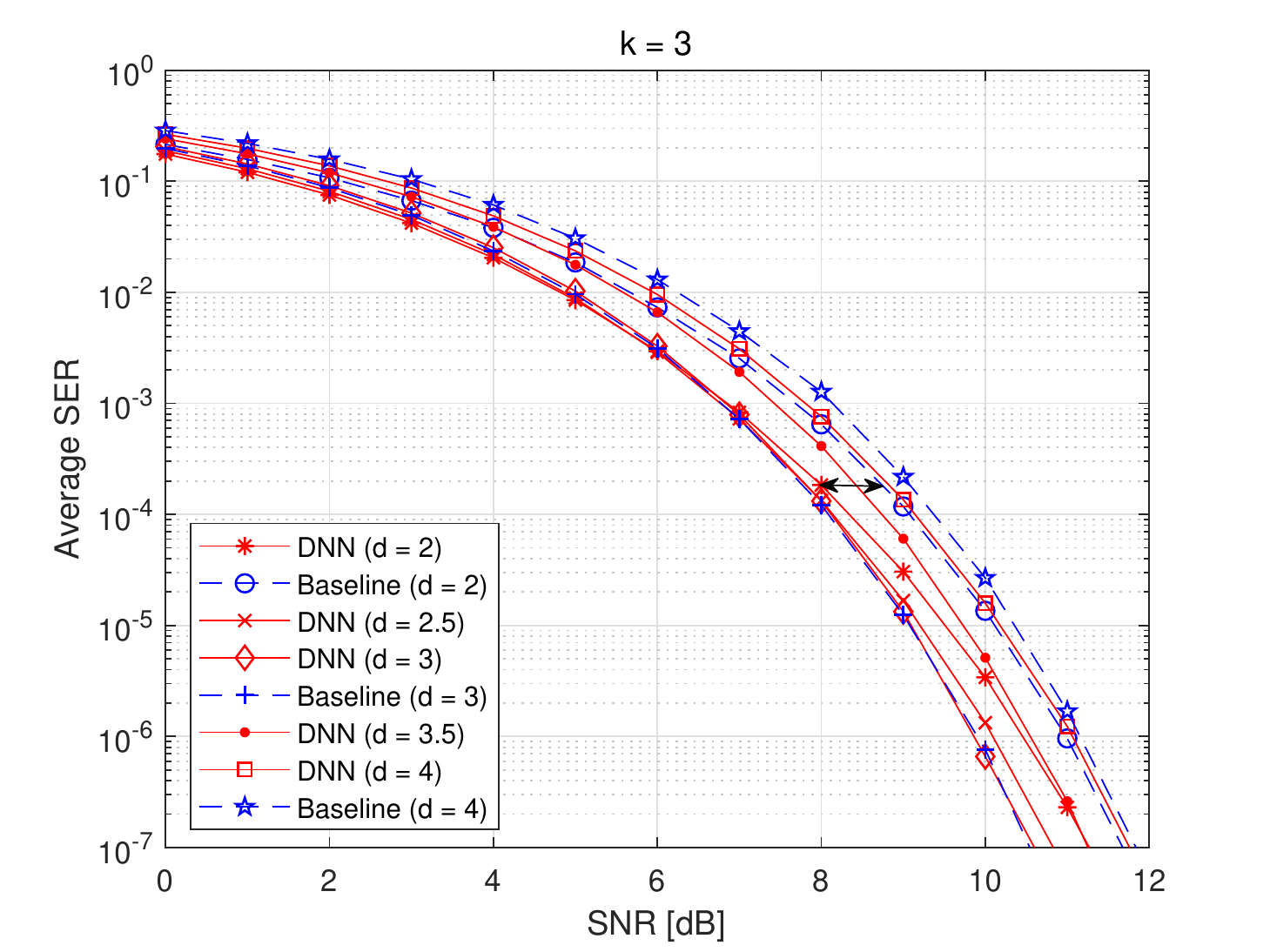}
    }
    \subfigure[$k=4$]{
        \includegraphics[width=.93\linewidth]{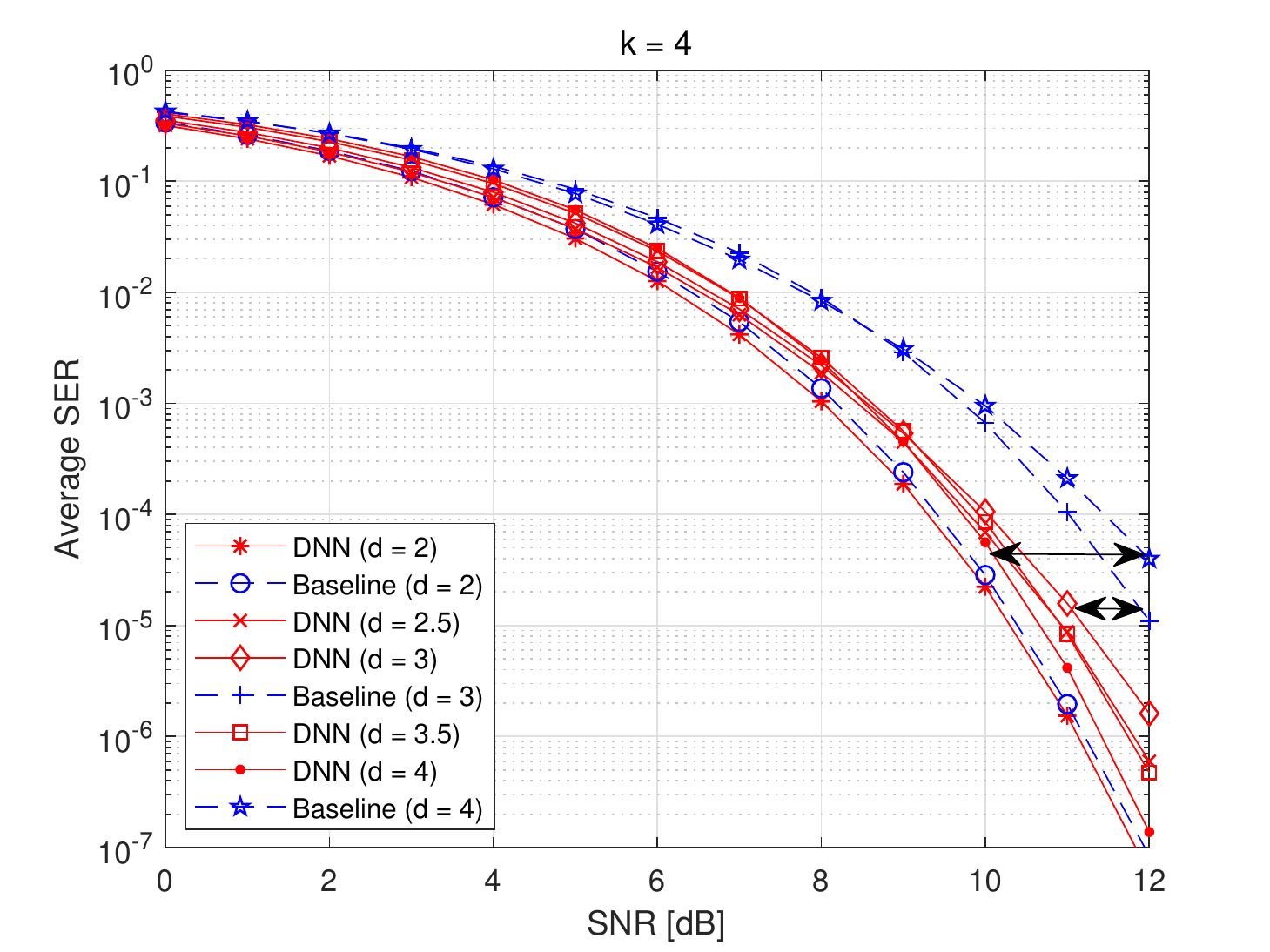}
    }
    \caption{Average SER performance as a function of SNR for $N=8$.}
    \label{fig:fig5}
\end{figure}

Fig. \ref{fig:fig5} shows the average SER performance of the trained DNN transceiver is evaluated over the test set by varying the SNR. For comparison, a VLC system in \cite{SZhao:16} is considered as a baseline, where the ML decoding rule is used for the symbol recovery. To further enhance the performance of the dimming control code in \cite{SZhao:16} we numerically find CWCs that achieve theoretical minimum Hamming distance profiles \cite{Ostergard:10}. Such CWCs can be searched by iteratively improving the solution for random choice of candidates. It is emphasized that, although the proposed DNN transceiver is trained at a certain SNR, it consistently works well in all SNR ranges regardless of the dimming constraint. If $k=2$ and $d=4$, the proposed DNN transceiver provides over $1\ \text{dB}$ gain over the baseline VLC system albeit exhibiting the same minimum Hamming distance performance. This implies that the encoding network learns the stochastic properties of the optical channel by itself and generates codebooks robust to the optical noise. By contrast, the CWC \cite{Ostergard:10} focuses only on the maximization of the minimum Hamming distance, which results in a degraded SER performance. Similar phenomena can be observed for $k=3$ and $d=2$ as well as $k=4$ and $d=3$. In addition, the proposed DNN transceiver outperforms the baseline VLC for $k=4$ and $d=4$ offering a $2\ \text{dB}$ gain as a higher minimum Hamming distance can be obtained with the learned semi-CWC shown in Table~\ref{table:table1}. It is also observed that the gain obtained by the proposed DNN transceiver grows for increasing $d$. As the dimming constraint $d$ increases (up to $d=\frac{N}{2}$), the number of possible candidates for the codebook $\mathcal{S}_{d}$ also increases, implying that the design degree of freedom for the problem in \eqref{eq:P1} rises. Hence, the gain of the proposed DNN-based optimization becomes large in a high $d$ regime.

\begin{figure}
\begin{center}
\includegraphics[width=.93\linewidth]{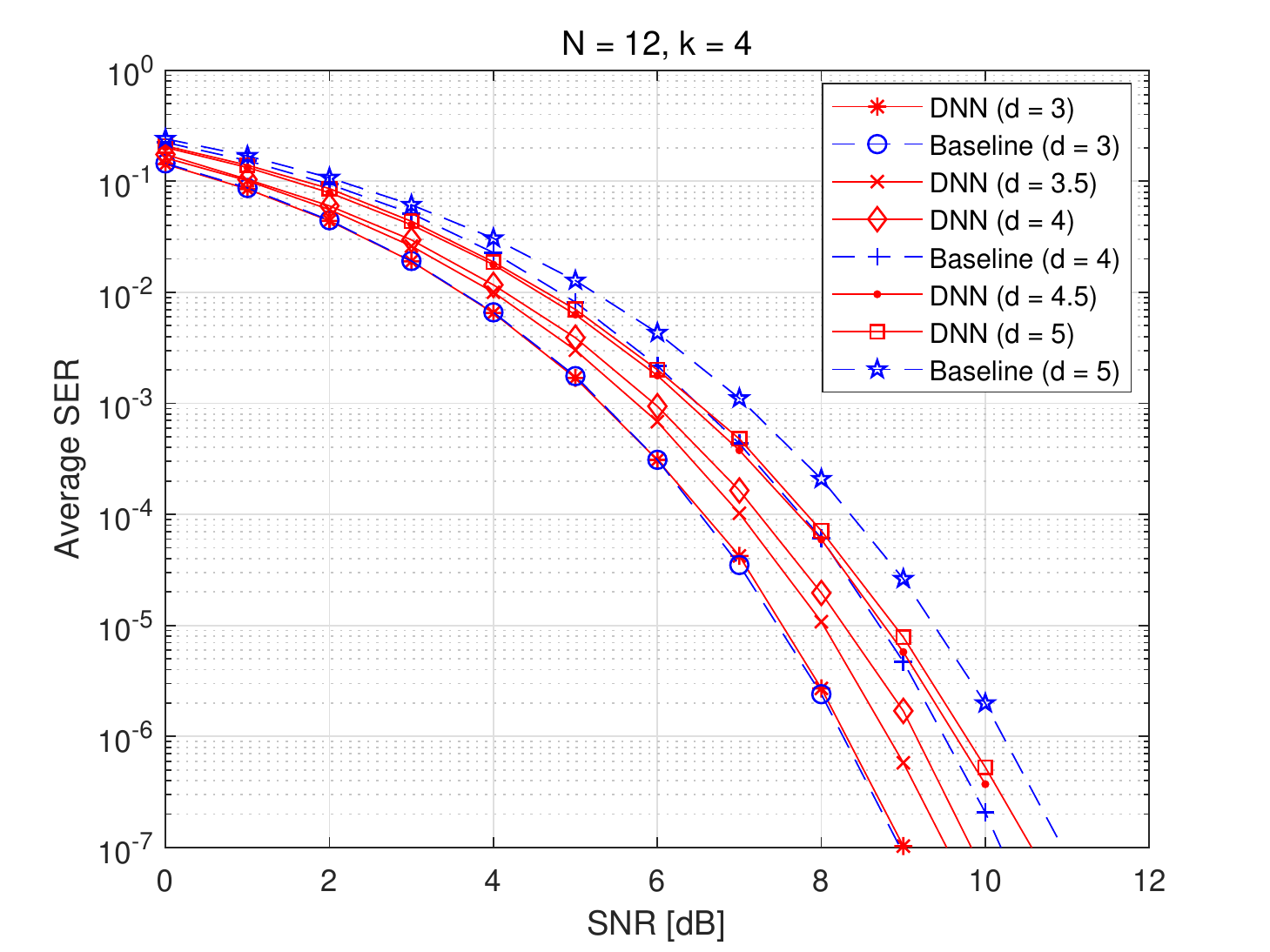}
\end{center}
\caption{Average SER performance as a function of SNR for $N=12$, $k=4$, and $\mathcal{D}=\{3,3.5,4,4.5,5\}$.}
\label{fig:fig6}
\end{figure}

Fig. \ref{fig:fig6} presents the average SER performance of the proposed DNN transceiver with $N=12$, $k=4$, and $\mathcal{D}=\{3,3.5,4,4.5,5\}$. We consider $L_{\text{E}}=L_{\text{D}}=4$ hidden layers both for the encoding and the decoding networks. For the encoding neural network, dimensions of the hidden layers are set to $24M^2$, $12M^2$, $12M^2$, and $6M^2$, respectively. Also, the hidden layers at the decoding neural network have the dimension $6M^2$, $12M^2$, $12M^2$, and $24M^2$, respectively. It is seen that the DNN technique can handle effectively multiple dimming constraints for $N=12$. We observe that, with the increased codeword length, the proposed DNN transceiver performs better than the baseline VLC system consistently. This indicates that the proposed DL framework for the design of the binary VLC systems can be scaled to an arbitrary codeword length.

\subsection{Nonlinearity and Varying Channel Setup}
The impact of the varying channel on the decoding performance is examined for a two-path ISI environment \cite{HWang:18} where the LED and the PD are deployed on the ceiling and the floor of a two-dimensional $3$ m-by-$3$ m room, respectively. Due to a wall located at $3$ m, there exists a LED-wall-PD reflection path incurring ISI to a LoS path from the LED to the PD. A commercial Kingbright blue T-1 3/4 LED \cite{Kbright:18} is considered with electro-to-opto transfer function $g(\mathbf{z})$ modeled as \eqref{eq:led_nl} with $K=4$, $\zeta=0.1$, $a_{1}=34.11$, $a_{2}=-29.99$, $a_{3}=6.999$, and $a_{4}=-0.1468$ \cite{WZhao:18}. The position of the LED is fixed at ($1.5$ m, $3$ m), while the PD suffering from the delayed ISI signal reflected by the wall is located at random on the floor as ($0$ m, $P$ m) with $P$ being the uniform random variable within $[0, 3]$. The LoS distance $D_{L-P}$ between the LED and the PD is given by $D_{L-P}=\sqrt{(1.5-P)^2+9}$ m, and the reflection distance $D_{L-W}$ from the LED to the wall and $D_{W-P}$ from the wall to the PD are $D_{L-W}=\sqrt{(\frac{4.5}{4.5-P})^2 + 2.25}$ m and $D_{W-P}=\sqrt{(3-P)^2+(3-\frac{4.5}{4.5-P})^2}$ m, respectively. According to \cite{HWang:18}, the received signal $\mathbf{r}$ in (\ref{eq:y}) for the ISI models can be written as
\begin{align}
    [\mathbf{r}]_{i}=&(1+\gamma(1-\Delta))[\mathbf{s}_{b,d}]_{i}+\gamma\Delta[\mathbf{s}_{b,d}]_{i-1}+[\mathbf{n}]_{i},\nonumber\\
    &\forall i=1,\cdots,N,\label{eq:rr}
\end{align}
where $[\mathbf{s}_{b,d}]_{0}\triangleq0$, $\gamma\triangleq\frac{D_{L-P}^{4}}{(D_{L-W}+D_{W-P})^{4}}$ and $\Delta\triangleq\frac{\tau_{d}}{T}$. Also, $T=10^{-8}\ \text{sec}$ \cite{HWang:18} is the bit time interval, $\tau_{d}\triangleq\frac{D_{L-W}+D_{W-P}}{c}$ represents the time delay of the ISI signal, and $c$ stands for the speed of the light. Then, the optical channel matrix $\mathbf{H}$ can be modeled as a Toeplitz matrix whose $(i,j)$-th element $[\mathbf{H}]_{ij}$ $(i,j=1,\cdots,N)$ is expressed as
\begin{align}\label{eq:ISI}
    [\mathbf{H}]_{ij}=\begin{cases}
    1+\gamma(1-\Delta), & \text{for}\ j=i,\\
    \gamma\Delta, & \text{for}\ j=i-1,\\
    0, & \text{else}.
  \end{cases}
\end{align}

\begin{figure}
\centering
    \subfigure[Linear LED]{
        \includegraphics[width=.93\linewidth]{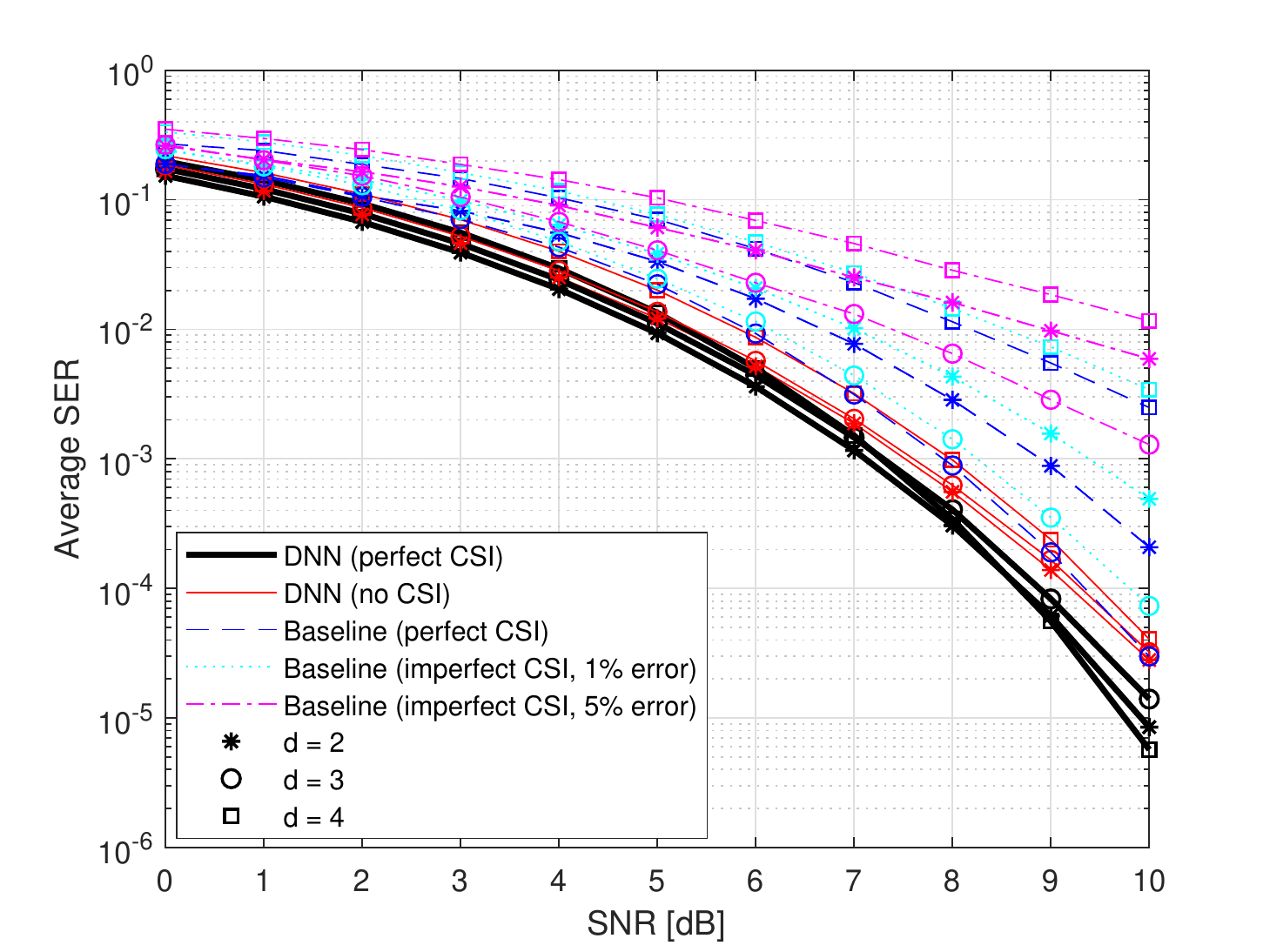}\label{fig:fig8a}
    }
    \subfigure[Nonlinear LED]{
        \includegraphics[width=.93\linewidth]{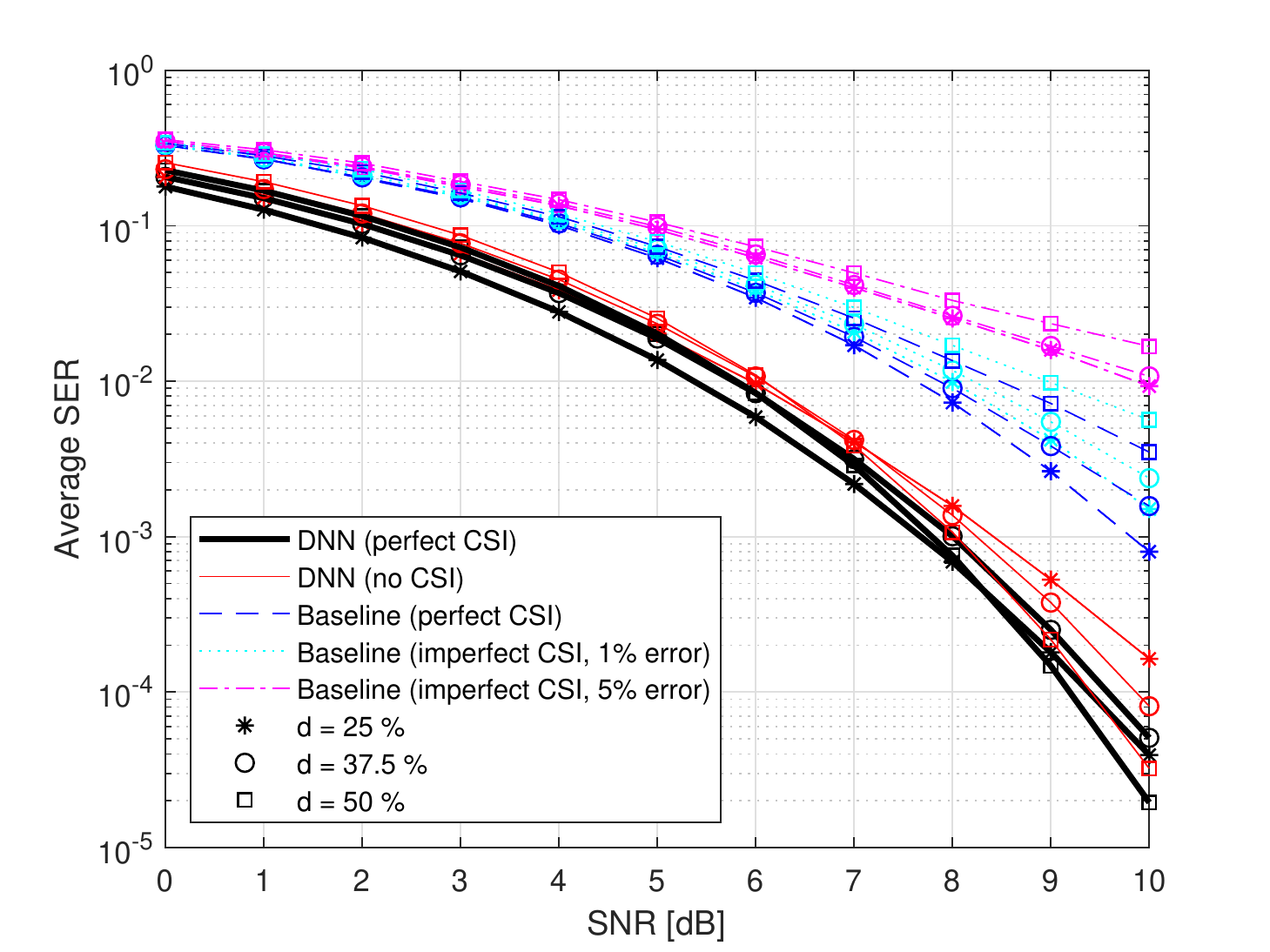}\label{fig:fig8b}
    }
    \caption{Average SER performance as a function of SNR for $N=8$ and $k=3$.}
    \label{fig:fig8}
\end{figure}

Fig. \ref{fig:fig8} shows the SER performance averaged over randomly varying ISI channel instances from the distribution specified by \eqref{eq:rr} and \eqref{eq:ISI} with $N=8$, $k=3$, and $\mathcal{D}=\{2,3,4\}$. To capture the statistical nature of the channel randomness, more complicated neural network architecture is considered with $L_{\text{E}}=L_{\text{D}}=5$ hidden layers both for the encoding and the decoding networks. The dimensions of the hidden layers of the encoding network are fixed as $32M^2$, $16M^2$, $8M^2$, $4M^2$, and $M^2$ respectively, and the decoding network is constructed with a reversed structure with the hidden layer dimensions $M^2$, $4M^2$, $8M^2$, $16M^2$, and $32M^2$, respectively. As discussed in Sec. \ref{sec:sec6}, two different design approaches are examined for the proposed DNN transceiver with perfect CSI and no CSI. Since the conventional ML decoder relies on the perfect knowledge of the CSI, for fair comparison, the performance of the baseline scheme is also evaluated with imperfect CSI case with different level of the CSI estimation error. First, Fig. \ref{fig:fig8a} plots the SER performance with the ideal linear LED. In the perfect CSI case, the proposed DNN transceiver outperforms the conventional VLC systems regardless of the dimming constraint. Interestingly, the DNN method without the CSI knowledge offers a significant gain over the baseline scheme with the perfect CSI. The superiority of the DNN-based blind detection mainly originates from the data-driven optimizing capability of an unsupervised training technique. Unlike the ML detection rule requiring the perfect CSI, the DNN receiver is trained to accurately decode the OOK codeword corrupted by a random ISI channel. During the training step, the DNN transceiver observes numerous received vectors and the corresponding channel distortions from the Lagrangian in \eqref{eq:Lag}. Thus, both encoding and decoding networks can extract useful statistic features of the ISI environment for end-to-end VLC systems. As a result, the encoding network can generate efficient OOK symbols for the ISI channel distribution in \eqref{eq:ISI}, and at the same time, the decoding network learns the blind detection strategy for the simulated environment. Consequently, the proposed DNN transceiver becomes robust to the random multipath effects with arbitrary located PDs.

Next, we present the average SER performance in Fig. \ref{fig:fig8b} by taking the LED nonlinearity into account. For the baseline scheme in the nonlinear LED case, we employ the CWC with the maximum Hamming distance is randomly found such that the nonlinear dimming \eqref{eq:dim_nl} is achieved. In addition, the electro-to-opto transfer function in \eqref{eq:led_nl} is included in the ML decoding. The proposed DNN methods exhibit a good SER performance regardless of the level of the CSI knowledge, whereas the performance of the baseline schemes is degraded in the nonlinear LED case. This verifies the effectiveness of the DNN-based VLC transceiver for handling LED nonlinearity as well as the varying channel effect.

\section{Conclusions and Future Works}\label{sec:sec7}
This paper has investigated a DL framework for dimmable OOK-modulated VLC systems. For universal support of arbitrary dimming requirement, the average SER minimization problem has been considered under multiple relaxed dimming constraints. To solve this combinatorial problem, we have proposed a DNN based VLC transceiver structure where modulation and demodulation of the end-to-end VLC system are addressed by encoding and decoding neural networks. The DNNs are trained such that the transmitted OOK symbol for any given dimming target is successfully recovered over the optical channel. To generate binary OOK signals, a novel stochastic binarization technique has been employed at the encoding network. An efficient constrained training algorithm has been presented for the binary DNN transceiver by optimizing the dual variables for universal dimming constraints along with the DNN parameters via a single training step. Numerical results have demonstrated the feasibility and the effectiveness of the trained DNN transceiver with arbitrary dimming support by handling the nonlinearity of LEDs and varying channel effects.

A major limitation of the DNN transceiver would be its robustness to the channel statistics unseen during the training step. This shortcoming mainly stems from an inherent nature of an offline DNN training strategy. One possible alternative is to train the DNN transceiver in an online manner, e.g., reinforcement learning technique \cite{Aoudia:19}, so as to adapt varying long-term channel statistics. Another future research direction is to extend a DNN structure accommodating long codewords. Recurrent neural networks \cite{YJiang:19} could be considered in long sequential OOK code design for practical VLC systems.

\nocite{*}
\bibliography{arXiv}
\bibliographystyle{ieeetr}

\end{document}